# Impact-Induced Differentiation in Icy Bodies

by

W. Brian Tonks[1], Elisabetta Pierazzo, and H. Jay Melosh
University of Arizona
Tucson, AZ 85721

1. Physics Department, Ricks College, Rexburg, ID 83460-0520

Submitted to Icarus
November, 1993
Revised July, 1997

**Length:**

31 manuscript pages

19 figures

2 tables

Why is this manuscript posted on arXive:

Brian Tonks wrote this paper in 1993 as part of his PhD research under my guidance (and with the assistance of then-graduate student Betty Pierazzo, now deceased). It was submitted to Icarus in 1993, reviewed and returned for revisions. After a long lapse of time, entailing new computations, Brian completed the revisions in 1997, just before he left to begin a postdoctoral year at AMES research lab under the guidance of J. B. Pollack. As often happens under these circumstances, his current work took precedence over old work and the revised manuscript was never resubmitted. Brian then received an offer of a faculty position in Physics at what was then Rexburg College (now Brigham Young University-Idaho at Rexburg), which he accepted. That position entailed considerable teaching but publication was not encouraged, and so this manuscript languished.

However, this work, which was current then, is still current and has garnered 6 citations (as an unpublished MS), some as recent as 2010-2013, according to Google Scholar. Revising it at the present time would be quite onerous and Brian, now a former department chair and considering retirement, is not currently interested in this task. However, with his oral permission, he accepted my suggestion (also suggested by Amy Barr, who cited this manuscript in a paper published in 2010) to post this on arXive so that other researchers may access this MS and its still-relevant results as exploration of the Jovian satellite system continues.

H. J. Melosh, 5 July 2016.

### *Abstract*

The icy bodies of the outer solar system have densities ranging from 1170 to about 2000 kg $m^{-3}$, implying that they are composed of a mixture of ice and a denser material, presumably silicates. It is thought that most of these bodies have differentiated into an icy mantle overlying a silicate core. If these objects grew in an impact environment similar to the late accretion environment of the terrestrial planets, large impacts may haveplayed a role in triggering the differentiation process. By the time icy objects grow to about the mass of Europa and silicate bodies grow to approximately lunar mass, a large high-speed impact can generate an intact melt region that allows the dense material to gravitationally segregate, forming a large density anomaly. If this anomaly generates sufficient differential stress, it rapidly segregates to the object's center, triggering whole body differentiation. We used an impact melting model based on the Hugoniot equations, the linear shock-particle velocity relationship, and the empirical relationship between shock pressure and distance coupled with a Monte Carlo simulation of the late accretion process to determine conditions under which large impacts trigger differentiation in icy bodies. In a gas-free environment, impacts of projectiles in the satellite's accretion zone have a small probability of triggering differentiation in surviving proto-satellites as small as Triton. The probability increases to 100% by the time surviving proto-satellites grow to the mass of Europa. The impact of heliocentric particles captured by the central planet also effectively triggers differentiation in the icy satellites. If the largest projectiles in the heliocentric distribution are comparable to the mass of the growing satellites, they trigger differentiation in all surviving proto-satellites by the time they grow to the mass of Callisto. However, Callisto and Ganymede have different probabilities of impact-induced differentiation if they captured a small fraction (0.01 and 0.5%) of the particles they accreted from a distribution of small (largest particles ~10% of the satellite's mass) heliocentric objects. Impact induced differentiation explains the apparent difference between these two objects in a plausible way.

## *Introduction*

Among the possible thermal effects of high-speed giant impacts is large-scale melting that under certain conditions can trigger whole planet differentiation. A class of objects that potentially undergo whole-body differentiation are the icy bodies of the outer solar system. With densities ranging from Mimas' 1170 (Morrison 1982) to Triton's 2064 kg $m^{-3}$ (McKinnon *et al.* 1995) , these objects are thought to be mixtures of ice and silicates, because water ice is the expected cosmochemically abundant volatile that sequesters the solar nebula's abundance of oxygen. If the density of ice is 1000 kg $m^{-3}$ and of silicates is 3000 kg $m^{-3}$, the rock:ice ratio (by mass) is about 1:10 for the 1170 kg $m^{-3}$ density object and 2:1 for Triton Because radioactive heat sources within the body's silicate portion are sufficiently abundant to cause eventual melting of ice, it is thought that most of these bodies have differentiated into an icy mantle and silicate core. However, large impacts may have triggered differentiation contemporaneously with accretion.

Current ideas about the formation of satellite systems were reviewed by Stevenson *et al.* (1986) . Conditions that existed during satellite formation are not well constrained. It is thought that the regular satellite systems of the outer planets formed in a manner analogous to the accretion of planets in the solar nebula. There are, however, major differences. Gas drag was probably important during satellite formation. Non-volatile components may not have been fully vaporized (Lunine *et al.* 1991). If they were, they would have condensed out of the nebula well before ices. Silicate dust grains may have then coagulated into relatively large (10's of km) objects before ice condensed. Such objects may have served as cores onto which ice condensed to form largely icy planetesimals that accreted later to form the icy satellites. More probable is coagulation of silicates into meter scale rocks that were incorporated approximately uniformly into icy planetesimals. It is also conceivable that silicates remained as flaky, grain-sized objects that formed an intimate mixture with ice to make up proto-satellites. It is thought that most of these bodies have differentiated into an ice mantle and silicate center in a manner analogous to the segregation of iron from silicates, although the lack of tectonics on and the general dark appearance of Callisto have been attributed to failure of that body to differentiate (Schubert *et al*. 1981), a proposal that was spectacularly confirmed by the recent Galileo encounters with Callisto (Anderson et al. 1997). Callisto's undifferentiated state stands in stark contrast to the evidence for magnetic fields in both Ganymede and Io (Anderson et al. 1996 a, b) which suggest that both objects have metallic nickel-iron cores as well as silicate mantles. Previously proposed explanations of Callisto's lack of differentiation have not been particularly successful (McKinnon and Parmentier 1986). Lunine and Stevenson (1982) required a very specific range of planetesimal strength to prevent Callisto's differentation and Friedson and Stevenson (1983) suggest that the slightly higher silicate content of Ganymede allowed radioactive heating to cause differentiation in that satellite but not in Callisto. Ganymede may have passed through an orbital resonace with Io and Europe and suffered a tidal pulse that Callisto escaped (Malhotra, 1991). In this paper we examine more closely the effect of high speed impacts on differentiation of icy bodies during the latter stages of their accretion. We will show that is plausible that impacts may have triggered whole-body differentiation in Ganymede but not in Callisto because of the differences in the mass and orbital distances of the two satellites. We examine the possibility that impacts triggered whole-body differentiation of the icy objects of the outer solar system by constructing Monte Carlo simulations of their formation.

We previously studied impact triggering of core formation within silicate bodies (Tonks and Melosh 1992) . Under plausible conditions, giant impacts may trigger core formation in silicate bodies as small as a lunar mass, and by the time silicate planets grow to the mass of Mars, the probability of impact-induced core formation is nearly 100%. As a result, the largest planetesimals that coalesced to form the terrestrial planets were already differentiated objects. Large impacts may also have characterized the formation of the icy bodies. Here, we estimate the conditions under which large impacts may trigger whole-body differentiation in these objects.

Besides the large difference in the condensation temperature between ice and silicates, there are other major differences between icy and silicate bodies. The shock pressure required to induce melting is much lower, around 10 GPa for ice (Ahrens and O'Keefe 1985) compared to ~100 GPa for silicates (Tonks and Melosh 1993) . This implies that lower speed projectiles generate significant melting. The lower melting shock pressure also implies that only ice and other volatile species melt–-silicates do not. The time scale for segregation from the melted ice is determined by the physical condition of the silicates. If they are small, flaky grains, the segregation time may be quite long. If they are large, roughly spherical rocks, they segregate much more rapidly.

Ice is inherently weaker than intact silicates. Rheological experiments (Kirby *et al.* 1987) show that cold ice can withstand differential stresses of about 1 kbar, compared to 10's of kbar for silicates. "Warm" ice undergoes ductile failure at lower deviatoric stresses. This is partially offset by the smaller density contrast between ice and rock (~2000 kg $m^{-3}$) than that of silicate-iron (~4500 kg $m^{-3}$). Furthermore, the planet's gravity is also weaker than that of a silicate planet of the same mass because its lower mean density requires a larger planetary radius. Thus, crater and melt excavation is more efficient on an icy body, which tends to suppress the formation of large intact melt regions.

Finally, most of the satellites of the outer solar system formed in the gravity fields of a large central planet. As a result, they were bombarded by a poplulation of heliocentric particles as well as the planetocentric population out of which most of their mass accreted. These heliocentric objects struck the satellites at higher speeds than particles in the satellite's own accretion zone, increasing both the chance of impact-induced core formation and catastrophic disruption. Accretion both with and without heliocentric particles is treated below.

The mechanism of impact-induced differentiation and the conditions under which it is viable are now discussed. Incorporation of this mechanism into a Monte Carlo simulation of satellite accretion is reviewed. The shock pressure required to induce melting in ice is derived. Results of Monte Carlo simulations of icy satellite growth are presented and implications for the differentiation of icy bodies are discussed.

## *Differentiation Model*

The mechanism of impact-induced differentiation and its numerical implementation were discussed in detail in Tonks and Melosh (1992, 1993). It is shown schematically in Figure 1. A high speed projectile collides with a primary body composed of a mixture of ice and silicates (or silicates and iron in the case of the terrestrial planets) containing inhomogeneities that are small compared to projectile dimensions. The primary is assumed to have a cold core of undifferentiated material, consistent with early growth in a low speed impact environment. The impact generates a shock wave that propagates through both objects. If the impact speed is above a certain minimum, the shock generates a melt volume comparable to the projectile's volume. The melt volume depends on the projectile's radius, density, impact speed and angle. Under conditions discussed below, the melt forms an intact melt pond, shown schematically in Figure 1a. We assume that only the ice melts because of the large difference in the melting points of silicates and ice. Silicates, being more dense, gravitationally segregate from the melt, accumulating into a relatively thick layer buried at the base of the melt region (Fig. 1b). This mass of dense material generates differential stresses that must be supported by the underlying material. Davies (1982) suggested that there is a limit to a material's long-term strength. Cold material lying at the center of the planet can withstand only a certain maximum differential stress before its ultimate strength is exceeded. Thus, if the differential stress is larger than a certain maximum, designated here as the "threshold stress", the dense layer rapidly flows into the primary's center. Note that this mechanism is valid only for the first differentiation event. After this mass anomaly flows to the center, the gravitational potential energy released heats the originally cold mantle and mobilizes smaller anomalies already present and permits anomalies created by future impacts to be added to the core relatively rapidly.

We coupled this model of differentiation with a Monte Carlo simulation of planetary growth to estimate if, and at what stage in a planet's growth history giant impacts trigger catastrophic whole-body differentiation in silicate bodies (Tonks and Melosh 1992, Tonks and Melosh 1993, Tonks 1993) . This work extends the model and applies it to icy satellites. The algorithm is briefly described below. Further details can be found in Tonks and Melosh (1992, 1993). Extensions of the model not included in the original work are discussed here in more detail.

The growth of a homogeneous icy body with an initial mass $m_0$ is modeled in an accretion environment dominated by large projectiles. Although the algorithm can use any digitized mass distribution, a cumulative mass distribution of the form

$$N(m) = Km^{-p} \tag{1}$$

is generated for calculational convenience, where $N(m)$ is the cumulative number of particles in the distribution with a mass greater than $m$, $K$ is a normalizing constant, and $p$ is the power of the distribution. The power $p$ was varied from 0.873 to 0.5 with a nominal value of 2/3 (Tonks and Melosh 1992, 1993). This range is consistent with a hierarchical accretion environment in which most of the system's mass is contained in the bodies at the distribution's large-mass end. We used distributions containing 10000 particles in the simulations below.

We did not additionally weight the distribution for the gravitational cross-section of the planet. To do so is somewhat a matter of taste, depending on whether the reader thinks that a power law, equation (1) is a better representation of the particles in space around the growing planet or of the population of objects that actually strikes the planet (which is our prejudice here--and neither the distribution of planetesimals nor of impacting objects in the early solar system is known with any accuracy). In veiw of the insensitivity of our results to the exact exponent we do not believe this is a major factor.

A particle is randomly selected from the mass distribution to strike the planet. Its impact angle is randomly selected from a $\cos i \sin i$ distribution (Shoemaker 1962) , where $i$ is the impact angle, measured from the vertical. Its impact velocity is the vector sum of the planet's escape velocity and the "approach velocity", $v_\infty$. The approach velocity is the particle's velocity relative to the planet a long distance from it. The distribution of approach velocities in the icy satellite's accretion zone was modeled using a Maxwell-Boltzmann distribution. Our experience with silicate bodies indicates that the exact functional form of the distribution is not particularly important as long as it has a non-zero probability of having both high and low speed projectiles (Tonks and Melosh 1992). The speed at the peak of the distribution ($v_{pk}$) is presumed to be related to the Safronov number (Safronov 1969) by

$$\theta = \frac{1}{2}\left(\frac{v_{esc}}{v_{pk}}\right)^2 \tag{2}$$

where θ is the Safronov number, and $v_{esc}$ is the escape velocity of the largest planet in the distribution We presumed that the Safronov number equals 0.3 and is approximately constant during the satellite's growth history (Wetherill and Stewart 1989). This value assumes a gas free accretion environment. Because the escape velocity of the largest planet in the distribution grows at each stage of the simulation, the distribution's peak velocity also increases as the planet grows. Distributions used for heliocentric particles are described later.

If the Safronov number remains constant throughout a planet's growth history, the relative gravitational cross-sectional area of all bodies in the distribution remains constant, and planetesimals grow at a rate that depends only on their relative geometrical cross-sections. This scenario is known as "hierarchical accretion". If the Safronov number increases, the increase in gravitational cross-section due to the planet's increased gravity grows faster than its decrease due to increased approach velocities. In this case, the growth rate of large planetesimals is enhanced over the growth rate of small planetesimals, causing a condition known as "runaway accretion". In extreme cases, runaway accretion results in only a few large planetesimals and most of the system's mass resides in small planetesimals (Greenberg *et al.* 1978, Greenberg 1989). In moderate cases, such as those outlined by Wetherill and Stewart (1989), runaway growth is limited and most of the system's mass resides in the largest bodies. The distribution is not continuous as implied by eq. (1), but the largest planetesimal is detached from the continuous distribution. We examined the effects of hierarchical accretion by assuming the growing object is the largest mass in the continuous distribution described by eq. (1) and limited runaway growth by assuming that the growing satellite initially has 10 times the mass of the largest planetesimal in the continuous distribution. Because our interest is in large impact accretion environments, we did not examine the extreme runaway growth case where most of the mass resides in small bodies and the probability of giant collisions is extremely small: If this were the case then impact-induced differentiation would clearly be of little importance.

Next we modeled the effect of the impact on the planet. A high velocity impact generates a shock wave that propagates through both the target and projectile. The shock does irreversible work on the material through which it passes, increasing its energy and entropy. Decompression from the shocked state is an isentropic process. The material's final state after the shock depends on the difference between the shock and ambient pressure, the material's initial temperature (and pressure), and its equation of state. If the material's shocked state entropy is as large as its entropy at the liquidus (for a given decompression pressure), the material melts upon decompression. If the shock state entropy is larger than the vaporization entropy, material vaporizes upon decompression. There is a gradation of states, including all degrees of partial melting (or in single component systems, all possible melt fractions of the two-phase system). The difference between the shock and ambient pressures required for complete melting is called the "melting shock pressure difference". In most cases, ambient pressure is negligible compared to the shock pressure, so most often the melting shock pressure difference is referred to as the melting shock pressure.

To find the complete melting shock pressure, the material's equation of state must be known. One of the most powerful equation of state techniques now available is the computer code known as ANEOS (Thompson and Lauson 1984) . ANEOS calculates the thermodynamic properties of a material by using approximations to the Helmholtz free energy in different regimes of validity, ensuring thermodynamic consistency. Development of a new ANEOS material requires 24 input parameters and the output must be matched to experimental data. Development of the ANEOS equation of state for water ice is presented in Appendix 1. We used this equation of state to calculate the Hugoniot curves for ice at three different initial temperatures, shown in Figure 2.

The complete melting shock pressure is determined from the Hugoniot curve. If the liquidus entropy is known, the complete melting shock pressure is determined by the intersection of the Hugoniot curve and the complete melt isentrope. This method is employed in Figure 2b to determine the complete melting shock pressures for ice initially at 95, 150, and 225 K, respectively and at an ambient pressure of 1 bar. ANEOS directly computes the liquidus entropy (about 3800 J·(kg K)$^{-1}$ for ice at a decompression pressure of 1 bar), giving complete melting shock pressures of about 11, 10, and 7 GPa for the three initial temperatures. The liquidus entropy is nearly constant and the Hugoniot is nearly vertical at pressures below ~$10^4$ bars, so the shock melting pressure is insensitive to decompression pressures below this value. The shock pressure associated with any degree of partial melting can be determined in the same way if the relationship between entropy and melt fraction is known. For pure water ice, this is a simple linear relationship because water ice is a single component material. The ice making up the icy bodies is much more complex thermodynamically because of the presence of minor components, such as ammonia ice. Ammonia's presence may affect the shock pressure required for low degrees of partial melting, but probably has little effect on the large degrees of partial melting required for the impact-induced differentiation mechanism.

Detailed numerical calculations show that shock wave's pressure decrease with distance from the impact site displays two distinct regimes of behavior. Close to the impact site, pressure declines slowly with distance. For calculational ease, the region of slowly declining pressure has been considered as constant, and the region is called the "isobaric core" (e.g., Croft 1982). Outside the isobaric core, pressure declines rapidly with distance. Numerical calculations indicate that pressure decline outside the isobaric core follows an approximate power law (O'Keefe and Ahrens 1977) . Measurements taken during underground nuclear tests indicate that the particle velocity, rather than pressure, declines by a more precise power law (Perret and Bass 1975) . Consequently, we presume that

$$u_p(r) = u_{po}\left(\frac{r_o}{r}\right)^n \tag{3}$$

where $u_p(r)$ is the particle velocity at a radial distance, $r$, outside the isobaric core, $r_o$ is the radius of the isobaric core, and $n$ is the power. The power $n$ was measured as 1.87 by Perret and Bass. It can be shown that $n$ = 2 if the shock is approximately constant in thickness and momentum is conserved (Melosh 1989, p. 62-63). More recent work indicates that $n$ might be as low as 1.6 in certain materials (V. Nemchimov, personal communication, 1993). We used the conservation of momentum value, $n$ = 2, in the calculations that follow. One calculation was performed using $n$ = 1.6 to study its effect.

The relationships between material parameters across the shock are given by the Hugoniot equations (see Melosh, 1989, p. 37 for a summary). These equations represent conservation of mass, momentum, and energy across the shock. We used the linear shock-particle velocity relationship to relate the shock velocity to the particle velocity. Although ice undergoes several high pressure phase transitions, each of which could be represented by a different set of linear shock-particle velocity parameters, we found that linear shock-particle velocity parameters of 1.8 km $s^{-1}$ and 1.3 give the best overall fit for ice (See Appendix 1).

The particle velocity inside the isobaric core is determined using the planar impact approximation (Melosh 1989, pp. 54-56). The radius of the isobaric core is determined by assuming that the projectile's kinetic energy partitioned into the target is uniformly distributed throughout the isobaric core. After finding the isobaric core radius, the Hugoniot equations can be combined with the linear shock-particle velocity relationship and eq. (3) to determine the shock pressure as a function of radial distance outside the isobaric core. If the ambient pressure is negligible compared to the shock pressure (true throughout most icy bodies), this equation can be solved for the radial distance corresponding to a shock pressure $P(r)$

$$r = r_o \left( \frac{2\chi v_i \cos i S_1}{\sqrt{S_o^2 + 4P(r)S_1/\rho} - S_o} \right)^{1/n} \tag{4}$$

where $\chi$ is the ratio of the particle velocity in the isobaric core (determined by the planar impact approximation) to the impact speed, $S_o$ and $S_1$ are the linear shock-particle velocity parameters, $v_i$ is the impact speed, and $\rho$ is the target material's density.

If $P(r)$ is the complete melting shock pressure, $r$ corresponds to the radius of the complete melt region. If $P(r)$ represents the shock pressure required to produce a specific degree of partial melt, $r$ is the radial limit of that degree of partial melt. Because shock pressure declines with distance, all material inside $r$ is shocked to at least pressure $P(r)$. Consequently, all material inside $r$ is melted to at least the degree of partial melt associated with $P(r)$. Once the geometry of the melt region is specified, eq. (4) can be used to determine the melt volume.

Two different melt region geometries have been used, and are shown schematically in Figure 3. We used a hemispherical isobaric core surrounded by a hemispherical melt region in our examination of silicate bodies (Tonks and Melosh 1992). Based on the numerical computations of O'Keefe and Ahrens (1977), Croft (1982) assumed that the isobaric core is a sphere buried to a depth equal to its radius. Its radius is a factor of $\sqrt{2}$ smaller than the hemispherical model's isobaric core and is surrounded by a spherical melt region, truncated by the satellite's surface. Melt not present due to the satellite's spherical surface is accounted for by multiplying the melt volume by a geometric correction factor, which is a function of the melt and planet radii. Recent numerical calculations (Pierazzo et al. 1997) show that the isobaric core (and hence the melt region) is more of a squashed sphere, with a shape somewhere between these two regular geometric shapes. These geometries can be thought of as end member models with reality lying somewhere in between.

We computed only the complete melt region's volume in most of the calculations reported here and all calculations previously reported for silicate bodies (Tonks and Melosh 1992). The amount of melt produced in the partial melt region can also be estimated using the above equations if the relationship between entropy and melt fraction is known. The partial melt region was broken into a 100 shells with the boundary of each shell corresponding to a 1% melt fraction difference. The Hugoniot curve for the specified initial temperature is used to find the shock pressure corresponding to the partial melt fraction bounding each shell using the procedure illustrated in Figure 2b. The radii of shell boundaries are then found using eq. (4) and each shell's volume is calculated. The volume equation applicable to the melt region's geometry, with its appropriate geometric correction are used. The melt volume in each shell is computed by multiplying the shell's average partial melt fraction by the shell's volume and is added to the complete melt region's volume. The effect of including the partial melt region is shown below.

This procedure is only used until the partial melt fraction is ≤ 50%. At about 50% melt fraction, the melt-crystal mush undergoes a major rheological transition because the crystal content approaches closest packing. As a result, the flow mechanism changes from viscous liquid flow to either porous flow or solid state creep. At this transition, there is a dramatic increase in the effective viscosity of the mush (Tonks and Melosh 1990, Solomatov and Stevenson 1993) . Segregating silicates can not readily flow through this highly viscous mush, which thus forms an effective boundary for silicate segregation. Melt in the interstitial region between crystals below the rheological boundary does not contribute its silicate content to the differentiation process.

Not all of the melt generated by the impact remains near the crater. A significant fraction is excavated during crater formation. This melt mixes with excavated solids and cools during its ballistic flight. Because it is removed from the intact melt region, it does not contribute to the differentiation process. The fraction of the melt volume that is not excavated from the crater is designated as the "retained melt fraction", $f$, and is given simply by

$$f = \frac{V_m - V_{\mathrm{exc}}}{V_m} \qquad (5)$$

where $V_m$ is the melt volume and $V_{\mathrm{exc}}$ is the excavated melt volume. The excavated volume is computed using the Maxwell $Z$-model (Maxwell 1977) , assuming that cratering flow can be described by $Z = 3$. The total melt volume is multiplied by the retained melt fraction to obtain the retained melt volume, which then segregates to form the density anomaly.

We next presume that silicates in the retained melt segregate to the base of the melt region. Because silicates may not have been vaporized in the planetary nebulae out of which the satellite systems formed, the physical condition of silicates in icy bodies before differentiation and their segregation time scale are not known. It is conceivable that silicate grains coagulated into large objects (10's to 100's of km) and served as cores for icy planetesimals. Silicates might have remained as small, flaky grains. More likely, silicate particle sizes were somewhere between these two extremes and were homogeneously incorporated into the icy matrix when ice condensed. A plausible lower limit to the silicate segregation time scale can be estimated using the following argument. Presume that the silicate particles are spherical and small enough that Stokes' law is applicable. The time scale for separation is then

$$t_s = \frac{9h\eta}{2\Delta\rho g a^2} \tag{6}$$

where $h$ is the depth of the fluid, $\eta$ is the liquid's viscosity (~$10^{-3}$ Pa·sec for liquid water), $\Delta\rho$ is the density difference between liquid water and silicates (~2500 kg $m^{-3}$), $g$ is the acceleration of gravity (~ 1 m $s^{-2}$ for these bodies), and $a$ is the particle radius. If silicates segregate through a 200 km melt region and are 1 cm in diameter, the segregation time scale is ~4 hours. If the silicate particles are 1 mm in diameter, the segregation time scale is ~400 hours. If silicates were meter scale rocks imbedded within an icy matrix, the segregation time scale is on the order of seconds to minutes (Stoke's law is no longer applicable). Thus, if an intact melt region forms, silicates plausibly segregate from liquid water on a time scale that is short compared to the probable cooling and isostatic adjustment time scales and the assumption made in this model is valid. If the less likely alternative of homogeneous accretion of small, flaky silicates grains is correct, the segregation time scale could be much longer because of the large drag experienced as they fall. If so, the assumption that silicates segregate from the melt before cooling and isostatic adjustment may not be valid.

Next, we compute the differential stress generated by the segregated silicates and determine whether catastrophic differentiation occurs. Rheological experiments performed by Kirby *et al.* (1987) show that cold ice can withstand a differential stress of about 1 kbar. "Warm" ice undergoes ductile failure at lower deviatoric stresses. Precise conditions for catastrophic failure are difficult to determine and almost certainly are not confined to a single value. If the icy body was warm before the impact, the mantle creeps around the segregated silicates, allowing them to eventually end up in the body's center. The time scale for this type of differentiation depends strongly on the body's initial temperature and on the stress generated by the density anomaly.

However, we are interested in conditions that permit the density anomaly's stress to overcome the strength threshold of the cold core of undifferentiated material formed during the low velocity stage of accretion, leading to the first catastrophic differentiation event. Most of the calculations performed presumed a threshold stress of 1 kbar rather than the 20 kbar we previousy adopted for silicate bodies, although 0.5 kbar was also used for one set of computations.

These individual elements are integrated into a Monte Carlo simulation of accretion in a large impact environment. The algorithm randomly selects a particle from the mass distribution to impact the primary. If the impact speed is large enough to induce melting to the edge of the isobaric core, the retained melt volume is calculated. The primary and projectiles are presumed to be homogenous bodies composed of 60% ice and 40% rock by mass. This ratio is appropriate for the satellite systems. For objects that formed from the solar nebula (such as Pluto and Triton) a 40:60 ice:rock ratio is more appropriate (Prinn 1993). Rock segregates to the base of the intact melt region and its differential stress is computed. If the differential stress is larger than the threshold stress or if melting to the primary's center occurs, a core is presumed to have formed. If not, the projectile's mass is added to the primary's mass, and a new projectile is selected from the distribution. The simulation continues until a core is formed, the primary is catastrophically disrupted, or the body grows to the mass of the largest icy bodies without differentiating. No attempt was made to track non-catastrophic impact, tidal, or radioactive heating. These sources may be sufficient to trigger differentiation and because they raise the temperature of the satellite, they decrease the size of the object that will undergo catastrophic differentiation. Thus, the numerical values calculated below are upper limits.

***Results-Accretion from the Satellite's Feeding Zone***

The outcome of these simulations are statistical in nature. A given growth history might result in accretion of a large, high speed projectile that triggers differentiation early on. On the other hand, it is conceivable that a satellite escapes the large, high speed impacts required for differentiation by this mechanism and grows to its present mass without differentiating. A large, high speed impact may also result in catastrophic disruption of the planet. Disruption was discussed in detail in Tonks and Melosh (1992). We computed the volume displaced from the crater using pi scaling theory (Schmidt and Housen 1987) . If this volume is larger than the planet's volume, we assumed the body was catastrophically disrupted. The simulation records the satellite's final mass, the result of its growth, and the number of particles that accrete to yield the result. Each curve shown is based on calculation of 1000 growth histories under identical conditions. Run conditions are summarized in the text and figure captions. The cumulative percentage of growth histories that yield catastrophic differentiation is plotted as a function of the planet's mass when differentiation occurs. These graphs can be thought of as the probability that a giant impact triggers differentiation by the time the satellite grows to mass $M$. It is assumed that all impacts generate an intact melt region. Conditions required for intact melt region formation are discussed in Appendix 2.

The present algorithm explicitly allows segregation to occur only from the retained melt, calculates both the hemispherical and truncated sphere models, and includes the geometric correction. Our earlier silicate body computations used only the hemispherical model and did not include the geometric correction and the retained melt fraction (Tonks and Melosh 1992).

Figures 4-8 result from the straightforward application of the computations performed for silicate objects to icy objects. They assume accretion of satellite accretion zone planetesimals in a gas free environment and gravitational stirring of planetesimals by the primary. Figure 4 results from presuming a melting pressure of 10 GPa, $\sigma_b$ = 1 kbar, and $p$ = 2/3. Both orderly and runaway accretion cases are shown. Additionally, the effect of using the different geometric models (HM = hemispherical model; SM = truncated sphere model) are plotted. Pluto-mass planets have a small, but non-zero probability of impact-triggered differentiation (if the truncated sphere model is accurate). By the time objects grow to the mass of Europa, impacts have triggered differentiation in nearly all of the planets that survive the accretion process. Only about 26-31% of icy bodies growing in an orderly accretion environment survive to a mass

where differentiation can occur. The rest are catastrophically disrupted. This is a much higher percentage than for silicate bodies, occurring mainly because an equal mass silicate body has a larger gravity by a factor of 1.5-2. Because the accreting objects are, on average, ten times smaller in the case of runaway accretion, nearly all satellites survive. The satellite's mass when impact induced differentiation begins to occur is nearly the same for both orderly and runaway accretion. It is determined primarily by the threshold stress and the geometric melt model assumed.

Because of the much greater tendency to cause disruption, the choice of the satellite's initial mass must be done more carefully than in the case of silicate bodies. The lower limit of initial mass is determined by computational time-the lower the initial mass, the more particles that must accrete and the longer the computation time. If the initial mass is chosen so that only a few particles accrete before differentiation, the satellite is artificially protected from earlier large impacts that might disrupt it. Orderly accretion computations run under the conditions of Figure 4 were started with an initial mass of $10^{22}$ kg, well below the mass at which differentiation begins. In this simulation, about 40% of the satellites formed cores (60% disruptions). Another simulation was performed starting with a 7.5 x $10^{21}$ kg initial satellite, which resulted in 71% disruptions. The $10^{22}$ kg beginning mass artificially protects the satellite from disruption. The simulations displayed in Fig. 4 began with a 5 x $10^{21}$ kg satellite and had 74% disruptions. An additional run began with a 2.5 x $10^{21}$ kg satellite and also had about 74% disruptions. Apparently a 5 x $10^{21}$ kg initial satellite is small enough not to artificially protect the satellite from disruption.

The smallest satellite that can potentially undergo differentiation by this mechanism is determined by the threshold stress, the satellite's gravity (a function of its density for a given mass), and secondarily by the melt model assumed. This is because the differential stress generated by the segregated silicates depends on gravity. On a larger satellite, a given size density anomaly generates a larger differential stress. The minimum mass that an icy body must have for the 1 kbar differential stress is about $10^{22}$ kg. The truncated sphere model predicts differentiation at a smaller satellite mass because geometric correction removes increasingly more melt volume from the hemispherical model as the projectile (hence the melt volume) grows larger.

Figure 5 shows the effect of changing the slope of the distribution. Only results from the hemispherical model are shown with a melting shock pressure of 10 GPa and a threshold stress of 1 kbar. There is a slight shift in the mass at which differentiation occurs to smaller bodies with $p$ = 1/2, but these are second order effects. Orderly accretion disrupts fewer satellites with $p$ = .873 because there are fewer large planetesimals.

Figure 6 shows the results of changing the melting pressure to 6 GPa (consistent with ice near its solidus) and changing the threshold stress. Only the truncated sphere model computations are shown and $p$ = 2/3. Note that the mass range covers two orders of magnitude rather than the one order of magnitude shown in the other figures. Decreasing the melting pressure decreases the smallest mass at which differentiation occurs. However, just as in silicate bodies, it is a second order effect. Lowering the threshold stress to 1/2 kbar shows the optimum conditions for impact-induced differentiation in icy bodies. Under these conditions, the truncated sphere model predicts that objects as small as 4 x $10^{21}$ kg (somewhat larger than the mass of Titania) have a finite probability that impacts induce differentiation. Impacts trigger differentiation in nearly all Pluto-Triton-mass bodies that survive accretion.

The effect of the Safronov number is shown in Figure 7. It is plausible that the Safronov number changed with time and is a function of the central body around which accretion is taking place. If gas is present, the effective Safronov number is larger. The Safronov number was set equal to 3.0 for these simulations. Two effects are obvious: The satellite's mass where the differentiation transition occurs increases and the number of disruptions decreases. If the Safronov number is larger, the approach velocity is a smaller fraction of the satellite's escape velocity. Consequently, the satellite must have a larger escape velocity to produce an impact speed large enough to generate the required large-scale melting. Because the impact speed is lower, the number of disruptions decreases substantially. About 25% of planets are still disrupted. This is still higher than for silicate bodies except when ~10 km $s^{-1}$ projectiles strike asteroid-mass objects. There is a significant difference in the probability of differentiation for the largest icy bodies. Neither Callisto nor Ganymede has a very high probability of forming a silicate core, although Ganymede's chance is slightly higher than Callisto's. In all other cases, both satellites have a 100% probability (since they survived the accretion process) of impacts triggering differentiation.

Figure 8 shows the effect of including the partial melt region (to 50% melt fraction). Conditions include a melting shock pressure of 10 GPa, threshold stress of 1 kbar, $p$ of 2/3, and use of the truncated sphere model. Including the partial melt region only slightly decreases the satellite's mass at which differentiation begins. This is because at the presumed initial temperature of 150 K, the partial melt region is narrow and contributes only a small melt volume to the differentiation process. If the body were near its solidus at the time of the giant impact, the partial melt region is more significant (Tonks and Melosh 1992; Tonks and Melosh 1993; Tonks 1993). In this situation, however, the time scale for a diapir sinking into the planet's center is quite short, and smaller impacts would generate smaller, rapidly sinking diapirs. Consequently, exclusion of the partial melt region in the earlier results does not significantly underestimate the mass at which giant impacts trigger differentiation.

Figure 8 also shows the result of changing the power in the particle velocity-distance relationship from 2 to 1.6 and including the partial melt region. An examination of the statistics of the two runs shows that two curves are virtually identical and are equivalent within the statistical uncertainty of the simulations. Differentiation usually occurs at impact speeds close to the minimum impact speed required to cause melting to the edge of the isobaric core (equal to 3.0 km $s^{-1}$ for a melt pressure of 10 GPa and a 60:40 ice/rock ratio by mass). The change to $n$ = 1.6 results only in expansion of the melt region beyond the isobaric core. Because of the rapid drop in shock pressure outside the isobaric core, the melt region outside the isobaric core contributes only a small fraction of the total melt volume in most cases. As a result, this change does not significantly increase the total melt volume, resulting in no statistically significant change in the differentiation process.

***Results-Accretion of Heliocentric Objects***

Most of the major satellites of the outer solar system are believed to have grown in the gravity fields of their central planets. It is not known at what stage during the central planet's growth the satellites were formed. It is almost certain, however, that the central planet accreted solid material from heliocentric space during satellite growth. Even in the present epoch, the giant planets accrete small quantities of heliocentric mass, mostly stray comets, but also some asteroidal material that is scattered outward by resonances. The satellites undoubtedly intercepted some of these objects during their growth. Because of their high speed, heliocentric bodies may also be important in triggering differentiation as well as in disrupting the satellites.

Incorporation of heliocentric material into the Monte Carlo simulation of impact-induced core formation is relatively simple in principle. A given (most likely small) fraction of the projectiles that the satellite accretes is assumed to be heliocentric. These objects are presumed to follow a power law distribution (eq. 1), so a second particle distribution was generated to represent them. The code selects a uniform deviate. If it is a number between 0 and the assumed heliocentric particle fraction, a particle is selected from the distribution representing the heliocentric particles. Otherwise, a particle is selected from the distribution representing projectiles from the satellite's accretion zone. If a heliocentric particle is chosen, its impact speed is determined as described below. The consequences of impact into the satellite are then determined as described above for planetocentric impactors.

The impact speed of a heliocentric particle can be roughly estimated as

$$v_i^2 = v_{\rm enc}^2 + v_{\rm sat}^2 + v_f^2 + v_{\rm esc}^2 = v_\infty^2 + v_{\rm esc}^2 \qquad (7)$$

where $v_{\rm enc}$ is the encounter velocity of heliocentric particles with respect to the central planet (i.e., the speed with which the projectile is traveling with respect to the planet a long distance from it), $v_{\rm sat}$ is the orbital velocity of the satellite, $v_f$ is the speed due to gravitational focusing, equal to the escape velocity of the central planet at the satellite's orbital distance, $v_\infty$ is the projectile's approach speed, and $v_{esc}$ is the escape velocity of the satellite (Lissauer *et al.* 1988) . The satellite's orbital speed is taken to be its circular orbit speed. The escape velocity of the satellite is generally negligible compared to the other speeds in the equation, although it is explicitly included in the computer simulation. The major difficulties in implementing the simulation are determining an appropriate encounter velocity distribution and the fraction of heliocentric particles accreted. Zahnle *et al.* (1991) considered speed distributions of a late volatile rich veneer of heliocentric objects that originated from three sources: Planetesimals in the Uranus-Neptune region of the solar system, Kuiper belt comets, and Oort cloud comets. They computed a range of encounter velocities of 4.4-21.8 km $s^{-1}$ for Uranus-Neptune planetesimals and a range from 5.4 to 31.5 km $s^{-1}$ for planetesimals from the Oort cloud (Kuiper belt comets had intermediate speeds) at Jupiter's distance from the Sun, based on the orbital properties of a uniformly distributed population of projectiles. Encounter velocities are lower at Saturn due to that planet's distance from the Sun. During satellite formation, it is likely that a larger fraction of the heliocentric population originated closer to the central planet than even the Uranus-Neptune region, although there is some probability (albeit small) that one of the high speed comets from the Oort cloud impacted the growing satellites. Hence, heliocentric particle encounter speeds range all the way from a low of 0 km $s^{-1}$ to a high given by the Oort cloud comets. The Maxwell-Boltzmann distribution was used to model this speed distribution. The speed of the distribution's peak ($v_{pk}$) was chosen such that a uniform deviate of 1.0 corresponded to the highest speed calculated by Zahnle *et al.* (1991) for Oort cloud comets. This yields a speed at the distribution's peak of 6.3 km $s^{-1}$ at Jupiter and 4.6 km $s^{-1}$ at Saturn. The mean encounter speed $\bar{v}$ of the Maxwell-Boltzmann distribution is related to the velocity at the distribution's peak by

$$\bar{v} = \sqrt{\frac{4}{\pi}}\, v_{pk} \tag{8}$$

(see Reif 1965, p. 269). This gives an average encounter speed at Jupiter of 7.7 km $s^{-1}$, comparable to the 8 km $s^{-1}$ mean encounter velocity for Jupiter estimated by Smith *et al.* (1979) .

This method gives an encounter velocity distribution that is dominated by objects near the distribution's peak without having to make assumptions about the heliocentric particle distribution during this stage of accretion. It allows for a small, but non-zero probability that objects encounter the central planet at both low speeds and the high Oort cloud speeds. The simulation is not overly sensitive to the encounter speed because the impact speed is largely dominated by the satellite's orbital speed and the gravitational focusing speed. Figure 9 shows the distribution of the approach speeds (which equals the impact speed minus the small satellite escape velocity) predicted by the numerical routine. The three curves identified as "Europa", "Ganymede", and "Callisto" have the same encounter velocity distribution, which is a function of the central planet's semimajor axis. The differences in their approach speeds are due almost exclusively to differences in the orbital distances of these satellites. Approach speeds are higher for satellites closer to Jupiter. Approach speeds on Titan are smaller than those of the Galilean satellites because of Titan's distance from Saturn, Saturn's smaller mass, and Saturn's position further away from the Sun, which decreases the mean encounter velocity.

The fraction of heliocentric particles accreted by the satellites during satellite formation is unknown. It is undoubtedly a complex function of the stage of the satellite's and central planet's growth, the position of the satellite in the central planet's gravitational field, and the approach speed. If the central planet is essentially full grown (or at least the core is fully formed), the fraction is probably quite low. The approach speed of heliocentric particles is large compared to the speed of particles in the satellite's own accretion zone. The probability of impact (gravitational cross section) decreases with an increased approach speed. As a result, objects in the satellite's own accretion zone are much more likely to strike the satellite than are heliocentric particles. These arguments suggest that the fraction of heliocentric objects that accrete to the satellite is quite low. Because approximately the same number of particles must cross through a smaller spherical surface, the inner satellites are subject to a higher heliocentric particle flux than are the outer satellites of a given central planet. The increase in flux, however, is partially compensated for by a decrease in the satellite's gravitational cross section that results from the higher approach speeds deeper in the planet's gravity well. The fraction of heliocentric particles was left as a free parameter and was varied from .0001 to 0.005. Table 1 lists the fraction of growth histories that resulted in core formation and disruption triggered by particles from the heliocentric source and the satellite's accretion zone. Figures 10 and 11 show the results of these computations. Run conditions include a melting pressure of 10 GPa, a mass distribution power of 2/3 (for both the satellite's accretion zone and the heliocentric particles), threshold stress of 1 kbar, runaway accretion for satellite accretion zone particles, a large heliocentric particle distribution (that is, the largest body in the distribution equals the satellite's mass at the beginning of the computation, which is allowed to grow as the satellite grows), and a power of 1.6 in the particle velocity-distance relationship.

Figure 10 shows the percentage of growth histories that result in differentiation (regardless of which set of particles triggered differentiation) as a function of the satellite's mass at differentiation. The heliocentric particle fraction is assumed to be 0.005. Included is the $n = 1.6$ plot from Figure 8 for comparison. The core formation transition begins for slightly smaller satellites than in the case of particles only form the satellite's accretion zone. Disruption, which is the difference between the maximum of the curves and 100%, is important in all cases. Disruption becomes more important deeper in the central planet's gravity well. There is only a 16% chance that a satellite growing in Europa's position will survive the accretion process. The mass at which core formation occurs in Ganymede, Callisto, and Titan all group near the same value. The mass at which core formation occurs in a Europa-position satellite is higher. This is probably because of the large number of disruptions, although it is not clear that this is significant. Heliocentric particles tend to trigger core formation in slightly smaller satellites than do the satellite accretion zone particles.

Although disruption is quite common in many of these evolutionary scenarios, the overall outcome of a “disruption” event may be different for a satellite around a major planet than for a planet in heliocentric orbit because the much smaller configuration space available to the fragments around a major planet may lead to quick reassembly of the satellite. Such disruption events are unlikely to cause differentiation, however, so that the reaccreted satellite will remain a homogeneous mixture until an impact large enough to produce substantial melting finally occurs. Thus, the size threshold at which impact-induced differentiation becomes probable is independent of the possibility of re-accretion.

Figure 11 shows the effect of the fraction of heliocentric particles that accrete during this stage of growth. Only a satellite in Europa's position was calculated because the effect is most extreme. The effect is similar for all satellites but less extreme for satellites that are not as deep in the central planet's gravity well. As the fraction of heliocentric particles increases, a larger fraction of satellites are disrupted. "Europa" only has a 16% chance of escaping disruption by the time a core is formed if the fraction of heliocentric particles is 0.005. The computation was truncated when a core formed. It is likely that continued bombardment after catastrophic core formation would result in an even smaller chance of survival. If the heliocentric fraction is 0.0001, "Europa" has about a 90% chance of survival, and if the heliocentric fraction is 0, "Europa" has around a 99% chance of survival. The satellite mass where core formation occurs also shifts to slightly larger objects. This is because the satellite accretion zone's particles provides a larger fraction of the impacts that trigger core formation. Because the impacts are at lower speeds, the planet's mass when core formation occurs tends to be larger. The plausible case of a heliocentric particle distribution in which the largest planetesimal is larger (or perhaps much larger) than the satellite was not modeled. The effect would be to increase the probability of disruption but the mass when catastrophic core formation occurs would not change significantly.

Figure 12 shows the effect of assuming that the heliocentric particle distribution is much smaller than the growing satellite. In this case, the heliocentric particle distribution was computed assuming that the satellite is 10 times the mass of the largest planetesimal in the heliocentric distribution (as well as in the satellite accretion zone distribution). This ratio is maintained during the satellite's growth. "Europa" was again chosen because the effect is the most extreme. Also shown is the "Europa" curve from Fig. 10. Disruptions decrease from nearly 84% to about 40%. However, the satellite mass at which core formation occurs remains about the same.

In a gaseous environment that may have existed during satellite formation, the approach velocities of particles in the satellite's accretion zone are significantly damped. This could be modeled by choosing a large Safronov number. As seen in Fig. 7, a Safronov number of 3.0 results in only a small chance of catastrophic differentiation even for the largest satellites. Consequently, only heliocentric particles have a strong probability of triggering differentiation in this case. This possibility was examined by revising the code to add the mass contributed by the satellite accretion zone objects but not allowing them to trigger differentiation. Only heliocentric particles were allowed to trigger differentiation (and disruption). Computations were done for both the large heliocentric and small heliocentric particle distributions. The large heliocentric particle distributions calculations showed that if the satellites survived, large impacts triggered differentiation in both Ganymede and Callisto. The fraction of satellites surviving the growth process is quite small. The probability that differentiation occurs does not reach its maximum until the satellites are nearly fully grown.

Small heliocentric particle distribution simulations were more interesting. A series of simulations were carried out in which the heliocentric fraction was varied from 0.005 to 0.0001. If the mass when core formation occurred exceeded the satellite's present mass, the satellite survived accretion without impact-induced differentiation. Figure 13 plots the percentage of growth histories resulting in impact-induced differentiation (i.e. the probability of differentiation) as a function of the heliocentric particle fraction. Because there were very few disruptions with the small heliocentric particle distribution, most of the difference between these curves and 100% is because the satellite grew to its present mass without differentiation occurring. It is clear that the heliocentric fraction has a pronounced effect on the probability of differentiation. At the smallest heliocentric fraction tested, the satellites had only 10-20% impact-induced differentiation. At the largest heliocentric fraction tested, the probability of impact-induced differentiation is much higher. Ganymede has a consistently greater probability of impact-induced differentiation than does Callisto. This is due both to its position deeper in Jupiter's gravity well and its larger mass. It is conceivable that under these conditions a heliocentric particle triggered differentiation in Ganymede but Callisto survived accretion without undergoing impact-induced differentiation. However, this does require rather specific conditions. It is more likely that both Ganymede and Callisto will be either differentiated or undifferentiated. Nevertheless, this process provides a plausible mechanism within the framework of satellite growth for Ganymede to be differentiated and Callisto to remain undifferentiated.

## *Conclusions*

Giant impacts form intact melt regions that can undergo gravitational separation, triggering whole-body differentiation by forming density anomalies with differential stresses in excess of the mantle's threshold stress. The smallest body for which the mechanism works is determined primarily by the threshold stress and the object's mean density. The mechanism is effective in silicate bodies larger than the Moon and in icy bodies perhaps as small as Triton. The mechanism depends on formation of an intact melt region, generated by giant impact on silicate planets as small as the Moon and on icy bodies the size of Europa. Consequently, impact-induced differentiation is probably effective in triggering whole-body differentiation in the larger known icy bodies. Heliocentric particles are also effective in triggering differentiation. If a small fraction of heliocentric particles struck the Galilean satellites growing in a gaseous environment, there is a difference in the probability of differentiation between Ganymede and Callisto because of the difference in these two satellite's orbital distances and mass. Impact-triggered differentiation by heliocentric particles could thus provide an explanation of the difference between these two objects within the framework of satellite growth as it is currently understood.

### *Appendix 1 Construction of an Equation of State for Ice*

Numerous equations of state for hydrocodes have been developed since the 1950's (Melosh, 1989, Appendix II). Most, however, are designed to model the behavior of compressed matter and have serious problems in representing two-phase regions where both solid and vapor are present.

The semianalytic equation of state developed by Thompson and Lauson (1984), in the form of a FORTRAN computer program called ANEOS, provides a full thermodynamic description of a given material. Thermodynamic consistency is attained by using analytic approximations to the Helmholtz free energy in various domains of density and temperature. From this, pressure, entropy, and specific internal energy, the heat capacity at constant volume, and the sound speed (needed to control stability of the hydrodynamic computation) are derived using standard thermodynamic relations. Liquid-vapor, solid-liquid, and solid-vapor phase transitions can be represented in a thermodynamically consistent manner, and the Gibbs phase rule is satisfied. Solid-solid phase transitions can also be treated, although in the current version of ANEOS they are incompatible with a detailed treatment of the solid-liquid region.

The properties of a given material are described in ANEOS by an array of 24 variables, which can be extended to 40 for a better description of mixed-phase material states, an array containing the atomic number of each element present in the material, and an array containing the mass fraction of each element. These arrays are already tabulated in the ANEOS equation of state library for some materials. None of these, however, provides a good description of water. Consequently, the ANEOS parameters had to be constucted from the beginning. Ahrens and O'Keefe (1995) published an analytic equation of state for water, but in this work we were able to incorporate more data than they had available, as well as assuring thermodynamic consistency through the phase transition.

Because the current version of ANEOS cannot treat the solid-solid and solid-liquid phase transitions in detail simultaneously, we chose to ignore the former in favor of the latter. This is because the low shock melting pressure of ice implies an almost certain melting/vaporization of ice as a result of an impact event. It is the solid-vapor and solid-liquid transitions that are important to the application addressed in this work. Another difficulty in the treatment of water is its anomalous behavior in the melting region at pressures below 2 kbar. The slope of the solid-liquid phase boundary changes sign from negative at $P \leq 2$ kbar to positive at $P > 2$ kbar. Such an abrupt change of slope is not expected by ANEOS and had to be avoided. This was accomplished by artificially interpolating the phase boundary above 2 kbar down to zero pressure, which effectively extends the liquid field to lower temperatures than exist in reality, with a freezing point of -37.5° C. As we show later, this change does not affect the final determination of the equation of state for water significantly in our application, because it involves only a minimal part of the pressure range.

Numerous experimental data for water shocked to high pressures and temperatures are available. The most famous in the literature are the data of Walsh and Rice (1957) that reach pressure of about 40 GPa, those of Mitchell and Nellis (1982) that span the range from 34 to 320 GPa, and Lyzenga *et al*. (1982) , who present some temperature data for shocked water. We adjusted the ANEOS input parameters to fit the phase boundary curves for solid-liquid and liquid-vapor transitions, and to obtain a Hugoniot curve that fits the experimental data cited above. The results are shown in Figures A1-A3. The values of the input parameters we derived are reported in Table A1.

Figure A1 shows the fit of the calculated Hugoniot to the experimental shock data. Figures A2 and A3 show the fit of the calculated phase-boundary curves to the experimental curves. The extension of the liquid region at pressures below 2 Kbar is evident in Figure A2. Figure A3 shows the liquid-vapor phase boundary plotted in pressure-entropy space calculated by ANEOS compared to the measured phase boundary (taken from the ASME Steam Tables). The liquid side of the phase boundary fits quite well until near the critical point, but the vapor side of the boundary is badly represented and the critical point is displaced. The reason for the displacement of the critical point and the misrepresentation of the vapor side of the boundary is because ANEOS treats the vapor phase as a mixture of monatomic gases, rather than a molecular gas. Thus, the vapor entropy predicted by ANEOS includes the rather large entropy of dissociation of the molecular gas $H_2O$ into a monatomic gas mixture of hydrogen and oxygen. Even so, this should not affect the outcome of the calculations performed with this equation of state. The correct liquid and solid entropies and the fraction of vapor in the decompressed material can then be found using the well known vapor phase diagram.

### *Appendix 2: Formation of an Intact Melt Region*

Experience with terrestrial craters shows that the retained melt is either incorporated into the breccia lens in simple craters or forms a relatively thin melt sheet in complex craters. The melt sheet of the ~100 km diameter crater Sudbury forms an annulus at least 2.5 km thick around the crater's center (Grieve *et al.* 1991) . Crater scaling relationships (e.g., Schmidt and Housen 1987) coupled with the melt model described above imply that the melt volume increases more rapidly than crater volume with increasing projectile radius (Melosh, 1989 (see Fig. 7.7), Tonks and Melosh 1992, Grieve and Cintala 1992). Even the largest known terrestrial impact craters are dominated by solids; most material both excavated and displaced from the crater is solid. The retained melt is spread into a thin sheet over the crater floor. During the modification phase in large craters, the floor may be uplifted, causing the melt sheet to pool into an annulus. As the projectile diameter increases, the melt region's radius becomes a larger fraction of the crater radius. Additionally, the retained melt fraction increases. This implies that the melt sheet becomes thicker and that melt becomes increasingly more important in determining crater modification. During a giant impact on a large planet, a large fraction of material both excavated from and displaced to form the crater is melt. In this case, crater collapse must be dominated by the low viscosity melt rather than acoustically fluidized solids. The majority of the retained melt forms an deep, intact melt pond. The formation of this melt pond is required for impact-induced differentiation.

Determining a criterion for intact melt region formation is a difficult task because the transition is continuous. Any criterion is a somewhat arbitrary choice. There is some pooling of melt in craters as "small" as Sudbury, although probably not enough to form the deep, intact melt pond required for differentiation. Crater collapse in this structure was dominated by the rheology of solid material. If the melt radius is equal to the crater radius, the crater is formed entirely within the melt region. In this case, it seems obvious that melt rheology completely dominates crater collapse and the retained melt must pool to form an intact melt pond. This, however, is too conservative a criterion. The dominance of melt in regulating crater collapse depends on the fraction of melt in the material displaced from the crater. The larger the liquid fraction, the lower the effective viscosity of the mixture and the more important the melt becomes in determining the modification flow regime. If the liquid fraction displaced to form the crater is small, melt is unimportant and forms a surface sheet, as in Sudbury. If the retained melt is >50% of the entire crater's volume, the fluid flow mechanism is viscous flow of a liquid rather than solid state flow or acoustically fluidized flow. Additionally, there is a spatial effect- the liquid is concentrated where streamlines are nearly radial. As the melt region becomes a larger fraction of the crater volume, the remaining displaced solids are located along the crater sides where the streamlines flow upward and outward. This concentrates solids near the surface and causes the bulk of the melt to remain as a body. Consequently, the retained melt probably does not have to equal even half of the crater's volume. Based on these considerations, we assume that if the retained melt volume is greater than 40-50% the crater's volume, an intact melt region forms.

$$fV_m{}^h = 0.4\text{-}0.5V_c \tag{A2.1}$$

where $V_m{}^h$ is the melt volume not corrected for the planet's spherical shape and $V_c$ is the transient crater's volume. Because the crater volume is not corrected for the planet's spherical shape, it must be compared to the uncorrected melt volume. The crater's volume is found by assuming a crater geometry. Craters in solids typically assume a parabolic form with a depth to diameter ratio between 1/4-1/3 for most solids. This could be somewhat different in a giant impact due to the effects of lithostatic pressure and the planet's finite surface. Lacking better information, it is assumed that the transient crater is a paraboloid with a depth to diameter ratio of 1/3.

Figure A2.1 plots the ratio of retained melt volume to the crater volume for both the hemispherical model and the truncated sphere model. Figure A2.1a is computed for silicate planets; Figure A2.1b is for icy bodies. An intact melt region should form after impact by a giant projectile when a silicate planet is somewhere between the mass of the Moon and Mercury, but only when an icy body is near the mass of the largest known icy bodies. This is because these bodies have weaker gravity for a given mass, allowing more efficient melt excavation.

Figure A2.2 shows the effect of the presumed density on the projectile and planet. The calculations on which Figure A2.1b is based used a density of the water ice-silicate mixture of 1230 kg $m^{-3}$. The actual density of the large bodies is higher. Callisto, Titan, and Ganymede have mean densities of around 1800-1900 kg $m^{-3}$. The smaller icy satellites of Jupiter and Saturn have densities in the range of 1250-1500 kg $m^{-3}$. For comparison, the density of pure ice at 258 K (915 g/$cm^{-3}$; Bakanova *et al.* 1975) was compared to an ice-silicate mixture with a density of 1500 kg $m^{-3}$. Increasing the planet and projectile's density has a significant effect on the retained melt fraction. There are two primary reasons for this effect. Increasing the density decreases the planet's radius for a given planet mass. Consequently, the planet's gravitational acceleration increases. This has no effect (for a given impact speed) on the melt region's radius, but decreases the crater size. Hence, melt excavation is not as efficient. Additionally, a higher density decreases the impact speed required for melting to the edge of the isobaric core, producing more melting for a given impact speed. The calculation for silicate planets presumed a density of 3320 kg $m^{-3}$. It may have been more appropriate to use the density of a mixture of 70% hydrated silicates and 30% iron with a density of approximately 4300 kg $m^{-3}$. These considerations indicate that giant impacts generate intact melt regions in bodies larger than lunar mass silicate bodies and on icy bodies larger than Europa. The mass of the body depends somewhat on the projectile/planet mass ratio, the melting pressure, and impact speed. It also depends on the precise density of the material. However, the densities are fixed within certain small ranges by cosmochemical considerations. Of all these variables, only the projectile/planet mass ratio ranges over several orders of magnitude and hence principally determines the mass at which intact melt regions are formed.

Figure A2.3 shows the effect of changing the projectile to satellite mass ratio on the melt/crater volume ratio. A projectile that contains 20% of the satellite's mass fulfills criterion (A2.1) on a satellite with about half the mass of a satellite that is struck by a projectile containing about 5% of its mass.

*Figure Captions*

Figure 1 Schematic diagram of the impact induced differentiation mechanism. A large planetesimal strikes an icy body containing about 40 weight percent of homogeneously distributed silicate grains. The impact creates an approximately hemispherical shaped melt pond (Fig. 1a). Due to the density contrast between silicates and liquid water, silicates settle to the base of the melt region, forming an intact body with an anomalously high density (Fig. 1b). If this density anomaly creates a differential stress large enough to overcome the threshold stress of the underlying mantle, it quickly moves to the planet's center, triggering whole-body differentiation.

Figure 2a Ice Hugoniots calculated by the ANEOS equation of state package for ice, developed in Appendix 1, shown in pressure-entropy space. Also shown is the complete melting isentrope. The Hugoniot curve is the locus of all possible shock states to which a material can be shocked from a given initial state. Because decompression from the shocked state is an isentropic process, the decompression curve is a vertical line in pressure-entropy space.

Figure 2b. Ice Hugoniots and the complete melting isentrope, showing a narrow region of Fig. 2a. Knowing the complete melting isentrope for ice, the complete melting shock pressure can be extracted by finding the intersection of the Hugoniot curve and the complete melting isentrope. The melting pressures are found to be 11.2, 9.8, and 6.9 GPa for ice with an initial temperature of 95, 150, and 225 K respectively.

Figure 3a. Schematic diagram of the hemispherical melt of the isobaric core and melt region. This presumes that the hemispherical isobaric core is centered at the surface surrounded by a melt region of radius $r_m$.

Figure 3b. Schematic diagram of the truncated sphere model of the isobaric core and melt region. The spherical isobaric core is presumed to be buried to a depth equal to its radius and is surrounded by a sphere of radius $r_m$, truncated at the planet's surface.

Figure 4. Results of the accretion simulations. Conditions include a melting pressure of 10 GPa, threshold stress of 1 kbar, $p$ = 2/3, and Safronov number of 0.3. HM specifies the hemispherical model; SM specifies the truncated sphere model. These results can be thought of as the probability that an impact triggers differentiation by the time the planet grows to mass $M$. Triton-massed planets have a non-zero probability of having impacts trigger differentiation. By the time planets grow to the mass of Europa, impacts trigger differentiation of all surviving objects. The truncated sphere model has both a higher retained melt fraction and more total melt due to the geometric correction. As a result, the truncated sphere model predicts differentiation in slightly smaller planets.

Figure 5. Effect of changing the slope of the planetesimal power law distribution. The distribution with $p$ = 1/2 has more large projectiles than the distribution with $p$ = 0.873. This is reflected in the slightly smaller planetary mass at which differentiation occurs and in the larger fraction of disruptions under orderly accretion. The impact-induced differentiation process is not strongly dependent on the precise value of $p$.

Figure 6. Effect of changing complete shock melting pressure and threshold stress. Under these conditions, Pluto and Triton-mass bodies have a significant probability of an impact triggering differentiation, but Titania-mass bodies do not. Because these are the optimum conditions for impact-induced differentiation, the mechanism is not viable for the majority of the smaller known icy bodies.

Figure 7. Effect of the Safronov number. Because the Safronov number determines the impact speed, its value has a major effect on impact-induced differentiation. Changing the Safronov number by an order of magnitude shifts the differentiation transition by nearly an order of magnitude. Orderly accretion with a higher Safronov number causes fewer disruptions than for low Safronov numbers. If the Safronov number were as high as 3.0 as assumed here, only the very largest icy bodies have a non zero (albeit small) probability of impact-induced differentiation.

Figure 8 Effect of including the partial melt region and changing the power in the particle velocity-distance relationship. Conditions include runaway accretion, a shock melting pressure of 10 GPa, $p$ = 2/3, threshold stress = 1 kbar, and use of the truncated sphere model. Inclusion of the partial melt region allows differentiation in slightly smaller objects, but is a second order effect. Changing the power in the particle velocity-distance relationship to 1.6 (and including the partial melt region) produces a simulation that is statistically indistinguishable from the $n = 2$ simulation.

Figure 9 Approach speed distributions computed by the numerical technique utilized in the computation. The difference in approach speeds between the Jovian satellites is due to gravitational focusing and the satellite's orbital speed. Titan experiences approach speeds lower those the large Jovian satellites.

Figure 10 Effect of including of heliocentric particles. The fraction of heliocentric particles is 0.005 in all cases. The difference between 100% and the maximum of the curve represents the probability of catastrophic disruption. The probability of disruption decreases with distance from Jupiter. This is because the encounter speed decreases. The encounter speed at Titan is lower than the encounter speed at Callisto, resulting in even a smaller chance of disruption.

Figure 11 Effect of the heliocentric fraction. A satellite in Europa's position was considered in all cases because the effect is most extreme. The heliocentric fraction has a large effect on the fraction of growth histories that result in disruption, but only a small effect on the mass of the satellite when differentiation occurs. The probability that differentiation is triggered by a heliocentric particle also decreases.

Figure 12 Effect of the size of the heliocentric population. Figures 10 and 11 are based on heliocentric projectile populations with the largest mass in the distribution equaling the satellite's initial mass ($5 \times 10^{21}$ kg). The computation was repeated for a satellite in Europa's position assuming a heliocentric distribution with the largest mass equal to 0.1 the initial mass of Europa. The heliocentric fraction was assumed to be 0.005. Satellite accretion zone bodies have a larger probability of triggering differentiation than do heliocentric particles. Since particles in the small heliocentric particle distribution are smaller on the average, the fraction of growth histories resulting in disruptions decreases.

Figure 13 Probability of impact-induced differentiation only by heliocentric particles as a function of heliocentric particle fraction. The largest mass in heliocentric particle distribution is presumed to be 0.1 the satellite's initial mass (i.e. the small heliocentric particle distribution). If the heliocentric fraction is very small, the probability of impact-induced differentiation is low. Satellites have an excellent chance of growing to their present mass without impacts triggering differentiation. As the fraction of heliocentric particles increases, the probability of impact - induced differentiation also grows. Ganymede has a larger probability of impact-induced differentiation than Callisto. Symbols represent heliocentric fractions where the computations were performed.

Figure A1.1 Comparison between the Hugoniot computed from ANEOS equation of state for water and experimental data. Shock vs. particle velocity plot; data from Mitchell and Nellis (1982) and Walsh and Rice (1957, Fig A1.1a). Temperature vs. pressure plot; data from Lyzenga *et al*. (1982, Fig A1.1b). Both plots show the excellent agreement of the ANEOS equation of state with the experimental data at high pressures and temperatures.

Figure A1.2 Phase diagram for water in temperature-pressure space. The solid line is the solid-liquid phase boundary. Since ANEOS cannot treat solid-solid and solid-liquid phase transitions simultaneously, solid-solid phase boundaries have not been taken into consideration. The extension of the liquid region at pressures below 2 kbar is evident.

Figure A1.3. Hugoniot curve and phase diagrams generated by ANEOS for water ice using the input parameters of Table A1 in pressure-entropy space. Also plotted is the well known liquid-vapor phase diagram (from the ASME Steam Tables with the liquid's entropy matched with the liquid entropy computed by ANEOS at 273 K and 1 bar). CP is the critical point. Note that the liquid side of the liquid-vapor phase diagram matches that calculated by ANEOS until pressures near the critical point. The vapor side of the phase boundary does not fit well. This is because ANEOS treats the vapor phase as a monatomic mixture rather than a molecular gas. Thus, the entropy calculated by ANEOS includes the large entropy of dissociation not included by the steam tables. The fraction of liquid to vapor can be computed by following a constant entropy line from the Hugoniot to the decompression pressure and using the lever rule with the experimentally determined phase boundaries.

Figure A2.1 The ratio of the retained melt volume to the crater volume for silicate bodies (Fig. A2.1a) and icy bodies (Fig. A2.1b) Assumed conditions include melt pressure 115 GPa, impact speed of 15 km $s^{-1}$, uncompressed density = 3320 kg $m^{-3}$, dunite linear shock parameters for silicate bodies, melt pressure of 10 GPa, impact speed of 5 km $s^{-1}$, uncompressed density of 1230 kg $m^{-3}$, and ice shock parameters for icy bodies. Both figures presume the projectile mass/planet mass is 0.1. The intact melt pond criterion is fulfilled on silicate planets about the mass of Mercury, but only on icy bodies larger than Europa under these conditions.

Figure A2.2. Effect of presumed densities on the calculation of the retained melt fraction and the formation of an intact melt region. This figure was constructed by assuming densities of 915 $kg{\cdot}m^{-3}$ (pure water ice at about 258 K, Bakanova *et al*. 1975) and 1500 $kg{\cdot}m^{-3}$ (50:50 mixture of ice and silicates by mass) using water linear shock parameters derived in Appendix 1. A higher density increases the object's gravity (by shrinking its radius), decreases the projectile's radius (hence decreasing both the crater and isobaric core radii), and decreases the minimum impact speed required for extensive melting. These effects result in objects about an order of magnitude smaller having the capability of developing an intact melt region.

Figure A2.3 Effect of the projectile/planet mass ratio on the retained melt volume/crater volume ratio in icy bodies. Conditions are the same as in Figure A2.1b. The impact of a projectile containing 20% of a planet's mass generates an intact melt region on a planet about half as massive as the impact of a projectile containing 5% of a planet's mass.

Table 1  Results of accretion with heliocentric particles

| Satellite Position | Fraction of Heliocentric Particles | % Growth Histories that had Cores Formed by Heliocentric Particles | % Growth Histories that had Cores Formed by Accretion Zone Particles | % Growth Histories Disrupted by Heliocentric Particles | % Growth Histories Disrupted by Accretion Zone Particles |
|---|---|---|---|---|---|
| Europa | 0.005 | 12.2 | 0.9 | 86.0 | 0.9 |
| Ganymede | 0.005 | 14.5 | 2.2 | 82.8 | 0.5 |
| Callisto | 0.005 | 23.6 | 3.5 | 72.4 | 0.5 |
| Titan | 0.005 | 35.4 | 6.4 | 57.7 | 0.5 |
| Europa | 0.0005 | 28.2 | 46.6 | 24.2 | 1.0 |
| Europa | 0.0001 | 84.2 | 9.1 | 6.3 | 0.4 |

Table A1 Parameters Used for the Equation of State of Water

| Parameter | Description |
|---|---|
| Nel = 2 | No. of elements in material (H, O) |
| EOS = 4 | EoS Type: Solid-gas with electronic terms and detailed treatment of liquid/vapor region |
| $\rho_o = 1.1$ g/cm$^3$ | Reference density |
| $T_o = 233.15$ K = -40 C | Reference temperature |
| $p_o = 0$ | Reference pressure |
| $S_o = 1.8 \times 10^5$ cm/s | Shock velocity (from linear Hugoniot shock-particle velocity) |
| $\Gamma_o = 0.3$ | Reference Grüneisen coefficient |
| $\theta_o = 533.34$ K = 249 C | Reference Debye temperature |
| $S_1 = 1.3$ | From linear Hugoniot shock-particle velocity equation |
| $E_{sep} = 6.25 \times 10^{10}$ ergs/g | Zero-T separation energy |
| $T_m = 235.6$ K = -37.5 C | Melting temperature |
| $c_{53} = c_{54} = 0$ | Parameters for low density modifications to move the critical point |
| $H_o = c_{41} = 0$ | Thermal conductivity parameters; if 0 not included |
| $\rho_{min} = 0.8\rho_o$ | Lowest allowed solid density |
| $D_1 = \ldots = D_5 = 0$ | Solid-solid phase transition parameters |
| $H_{fus} = 2.3 \times 10^9$ ergs/g | Heat of fusion |
| $\rho_{liq} = 1$ g/cm$^3$ | Density of liquid at the melting point |
| $U_p = L_o = 0$ | Limits related to the cold compression relation for expanded states |
| $\alpha = 0.5$ | Parameter related to liquid EoS correction to match $p_{vap}$ data and boiling points |
| $\beta = 0.95$ | Same as above |
| $\gamma = 0.99$ | Same as above |
| $c_{60} = 0.4$ | Interpolation parameter in Grüneisen coefficient model |
| $c_{61} = 0$ | Same as above |
| $c_{62} = 0.3$ | Interpolation parameter in free energy expression |
| Flag = 1 | Ionization model: 0 = Saha; 1 = Thomas-Fermi |
| $E_{shift} = S_{shift} = 0$ | Shift energy and entropy for reactive chemistry modeling |
| f(H) = 2/3 | Fraction of H in molecule |
| f(O) = 1/3 | Fraction of O in molecule |

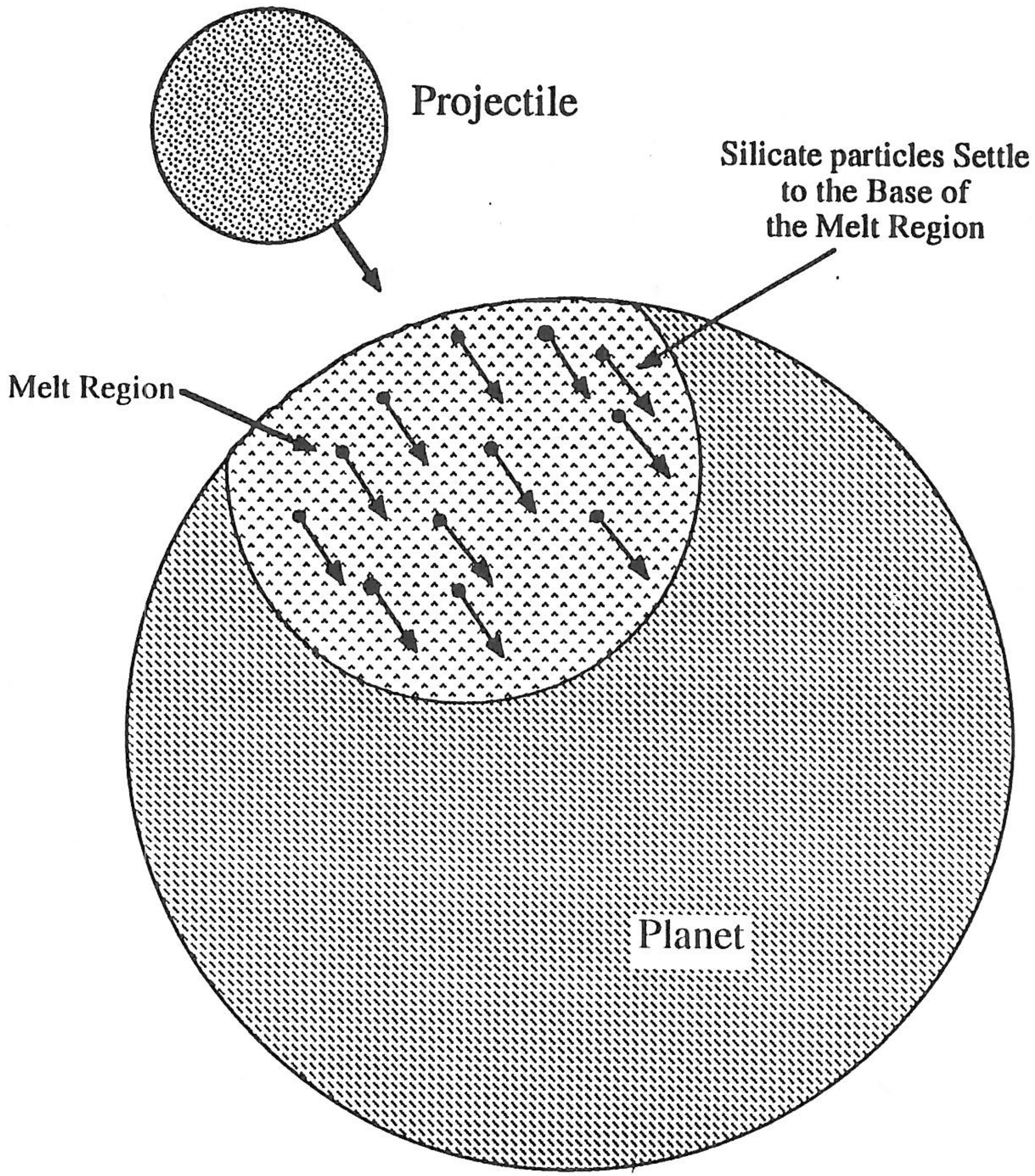
Projectile
Silicate particles Settle
to the Base of
the Melt Region
Melt Region
Planet

Fig 1a

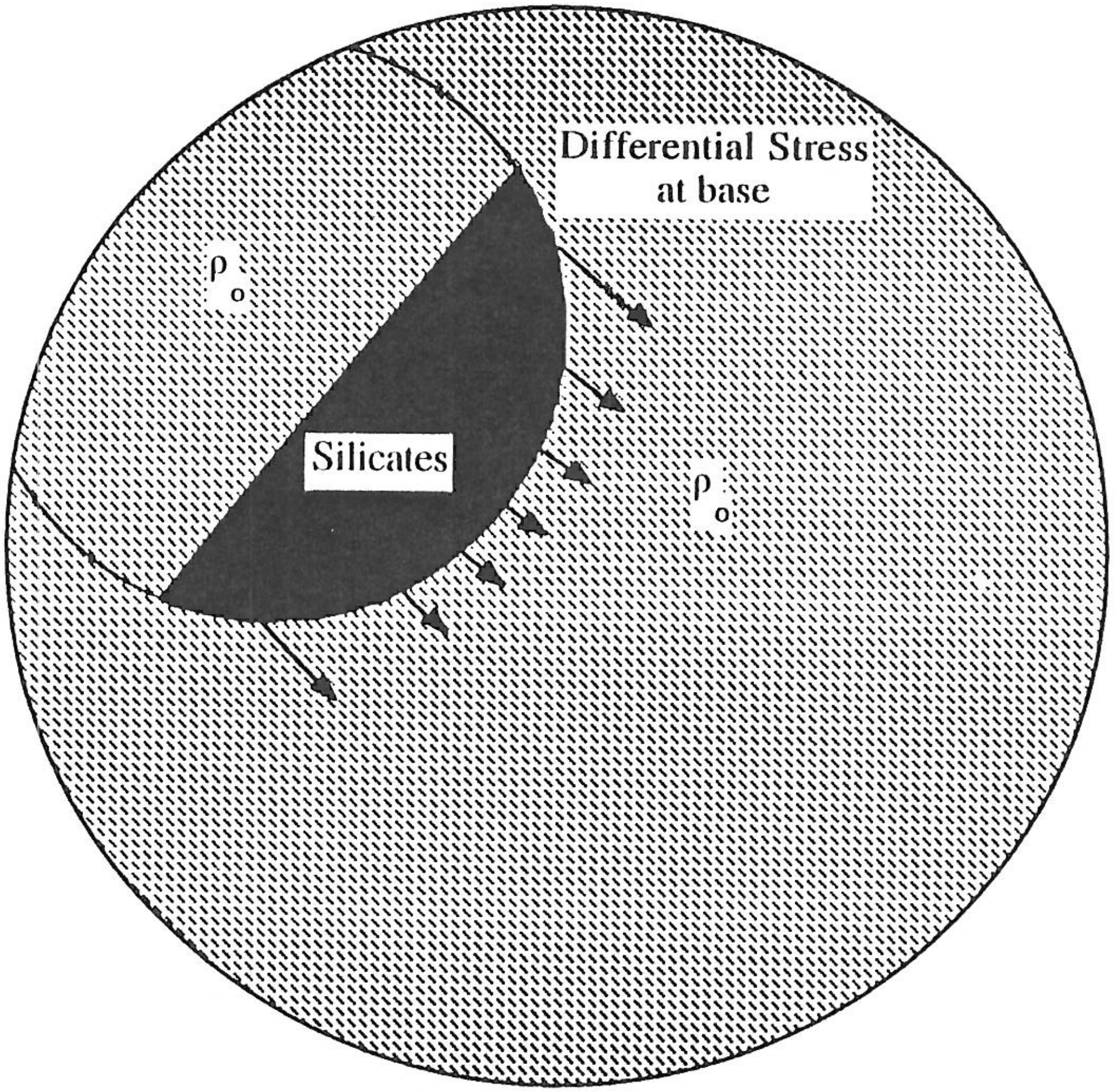

Fig 1b

Fig 29

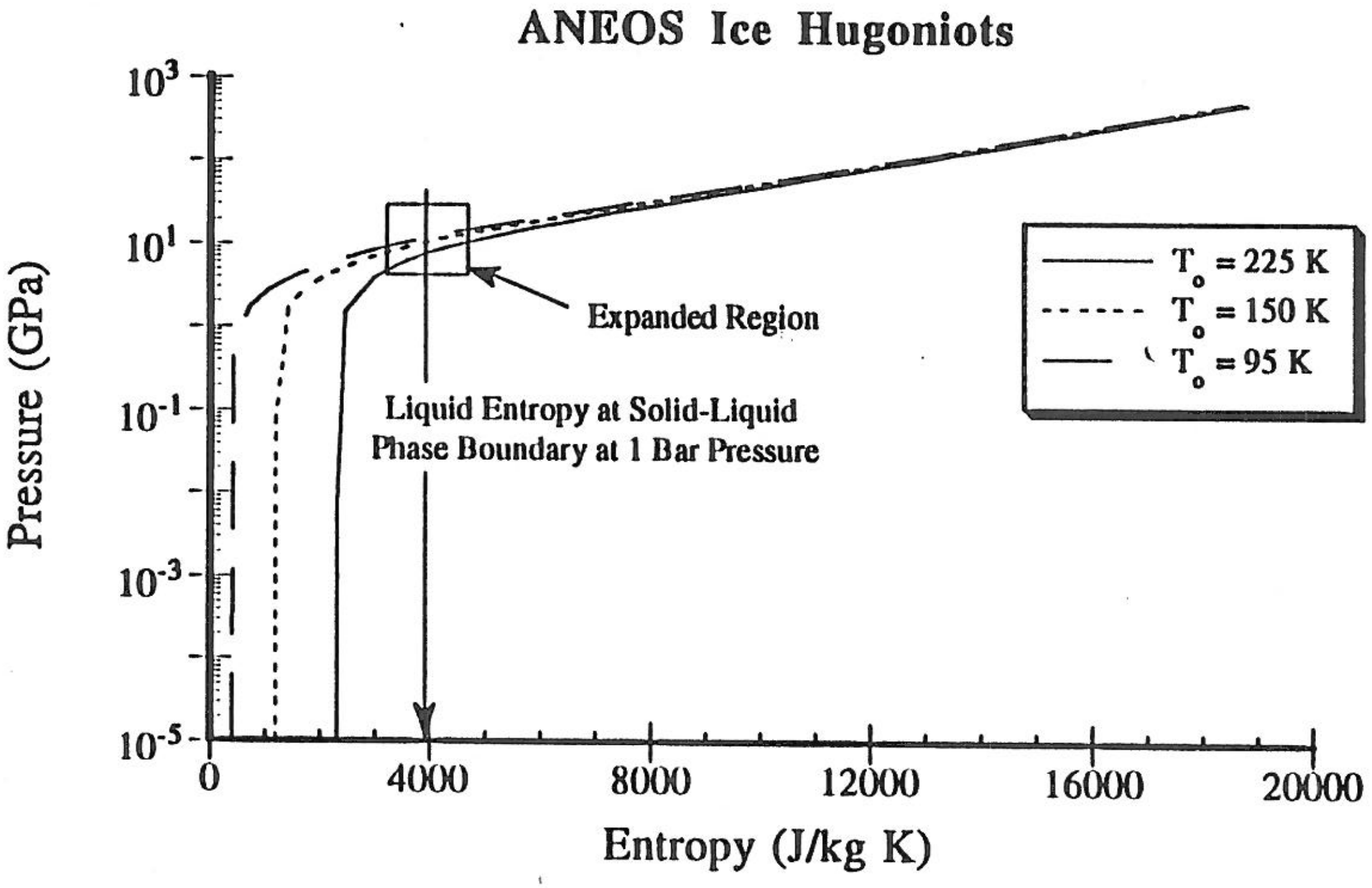

ANEOS Ice Hugoniots
Pressure (GPa)
10^3
10^1
10^-1
10^-3
10^-5
0
4000
8000
12000
16000
20000
Entropy (J/kg K)
Expanded Region
Liquid Entropy at Solid-Liquid
Phase Boundary at 1 Bar Pressure
T_o = 225 K
T_o = 150 K
T_o = 95 K

Fig 2b

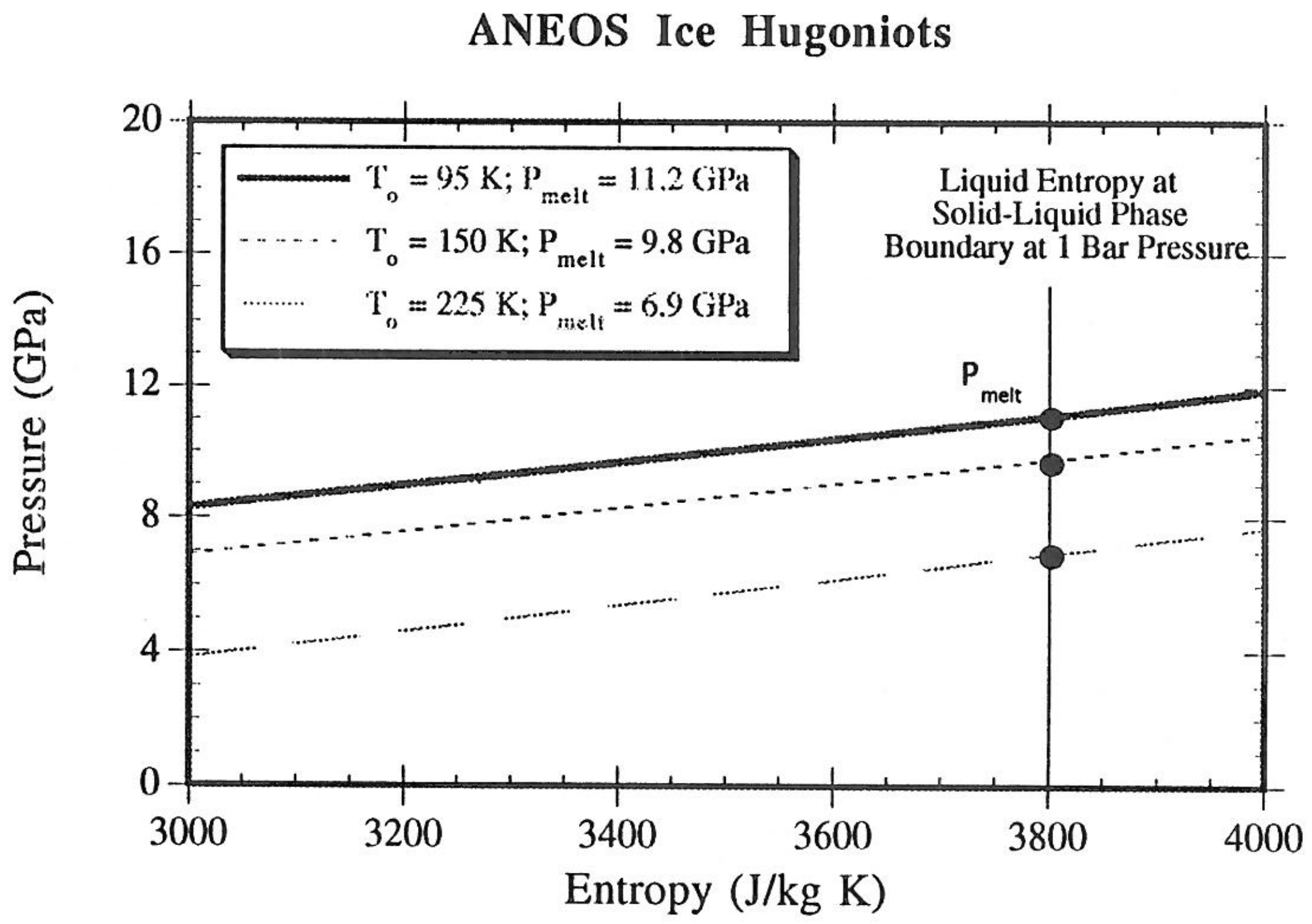

ANEOS Ice Hugoniots
$T_o$ = 95 K; $P_{melt}$ = 11.2 GPa
$T_o$ = 150 K; $P_{melt}$ = 9.8 GPa
$T_o$ = 225 K; $P_{melt}$ = 6.9 GPa
Liquid Entropy at Solid-Liquid Phase Boundary at 1 Bar Pressure
$P_{melt}$
Pressure (GPa)
0
4
8
12
16
20
3000
3200
3400
3600
3800
4000
Entropy (J/kg K)

Fig 3a

## Hemispherical Model

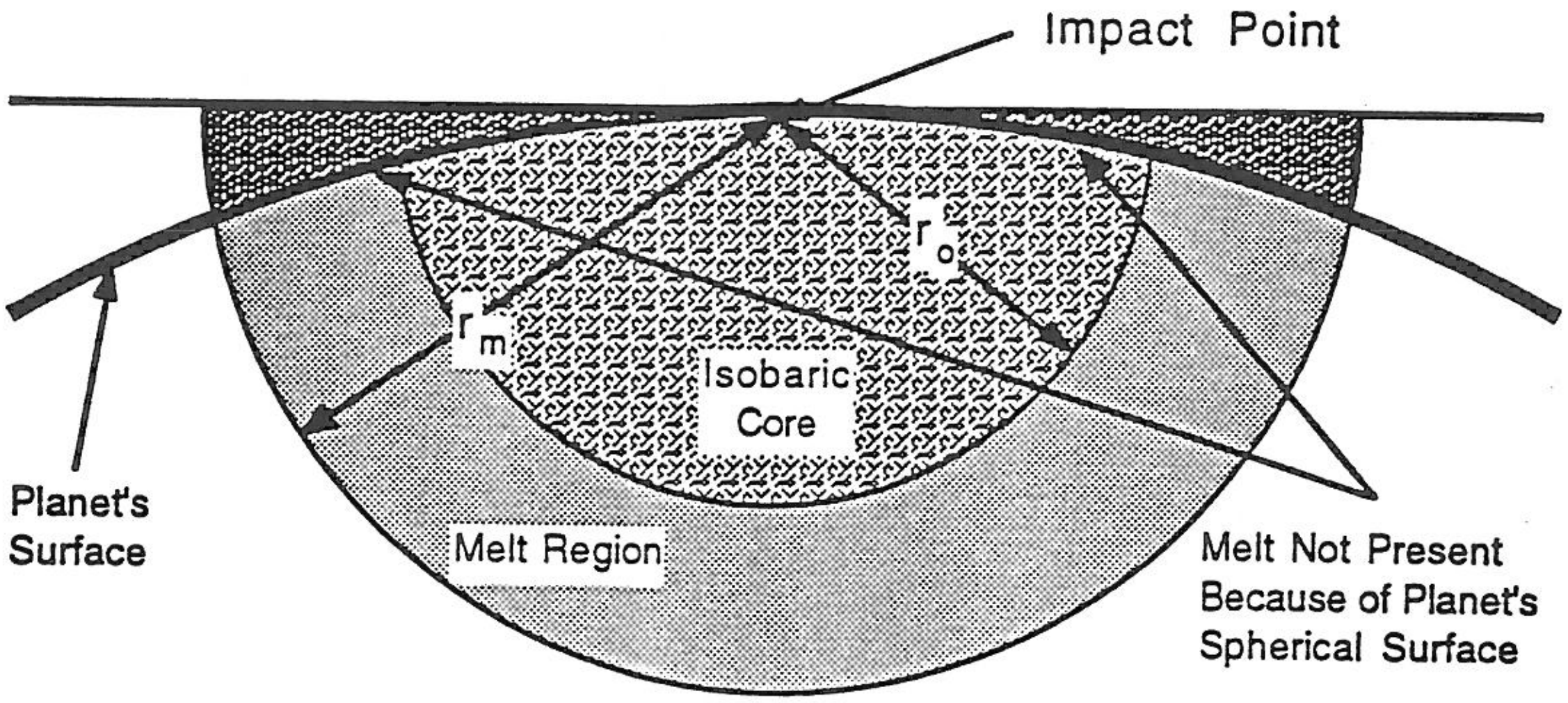

Fig 3b

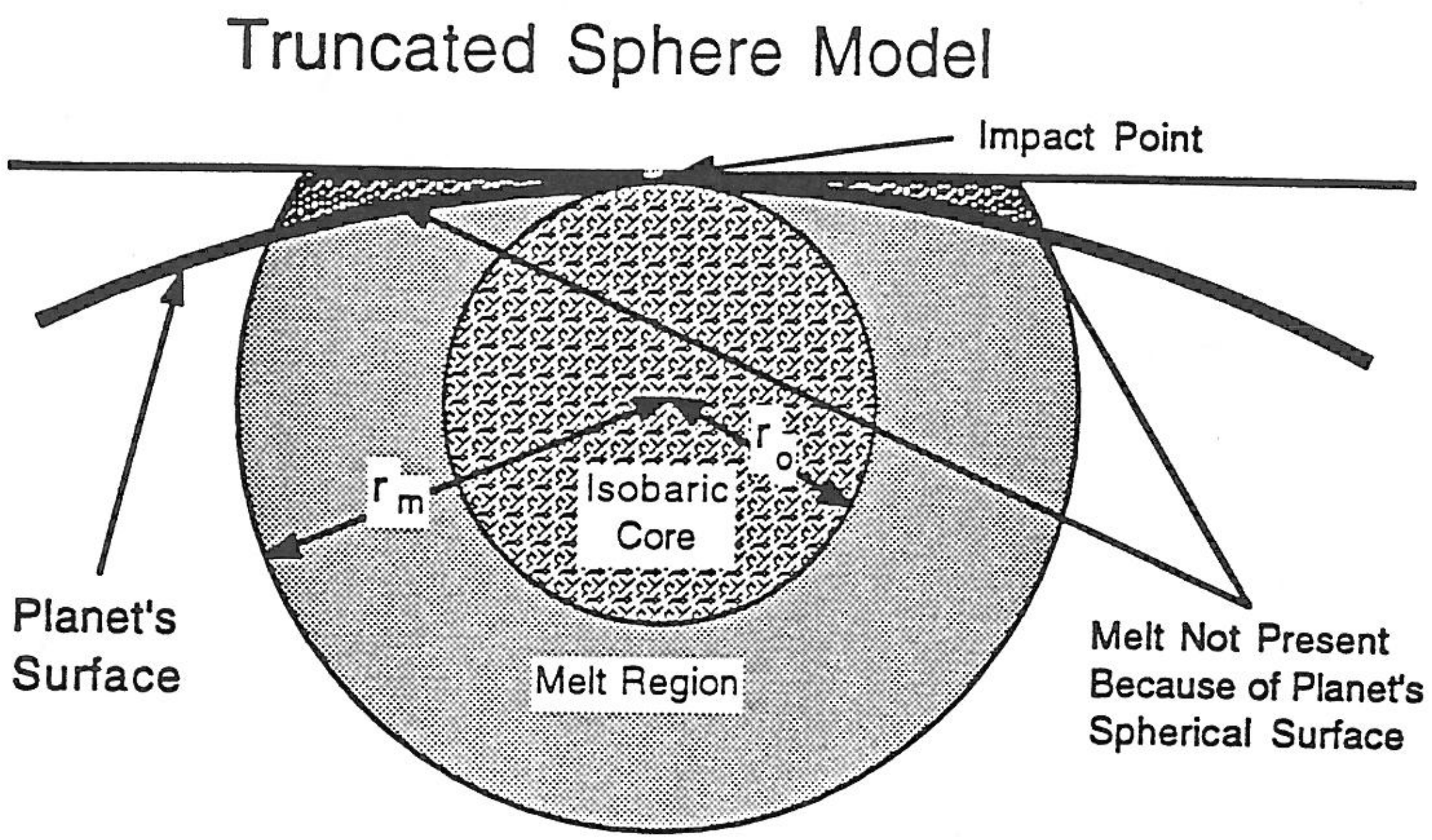
Truncated Sphere Model
Impact Point
$r_m$
$r_o$
Isobaric Core
Planet's Surface
Melt Region
Melt Not Present Because of Planet's Spherical Surface

Fig 4

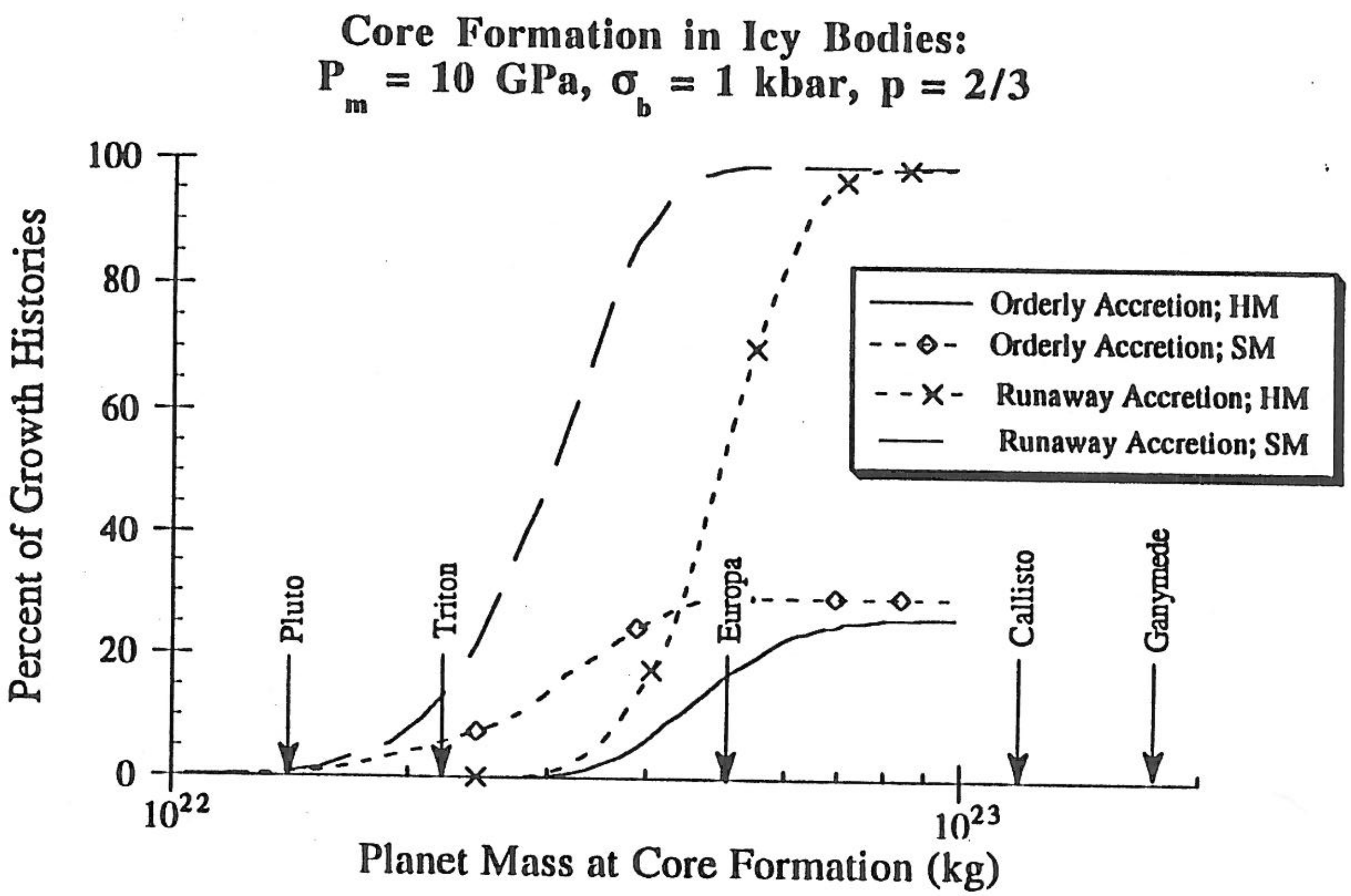

Core Formation in Icy Bodies:
$P_m$ = 10 GPa, $\sigma_b$ = 1 kbar, p = 2/3
Percent of Growth Histories
100
80
60
40
20
0
$10^{22}$
$10^{23}$
Planet Mass at Core Formation (kg)
Pluto
Triton
Europa
Callisto
Ganymede
Orderly Accretion; HM
Orderly Accretion; SM
Runaway Accretion; HM
Runaway Accretion; SM

Fig 5

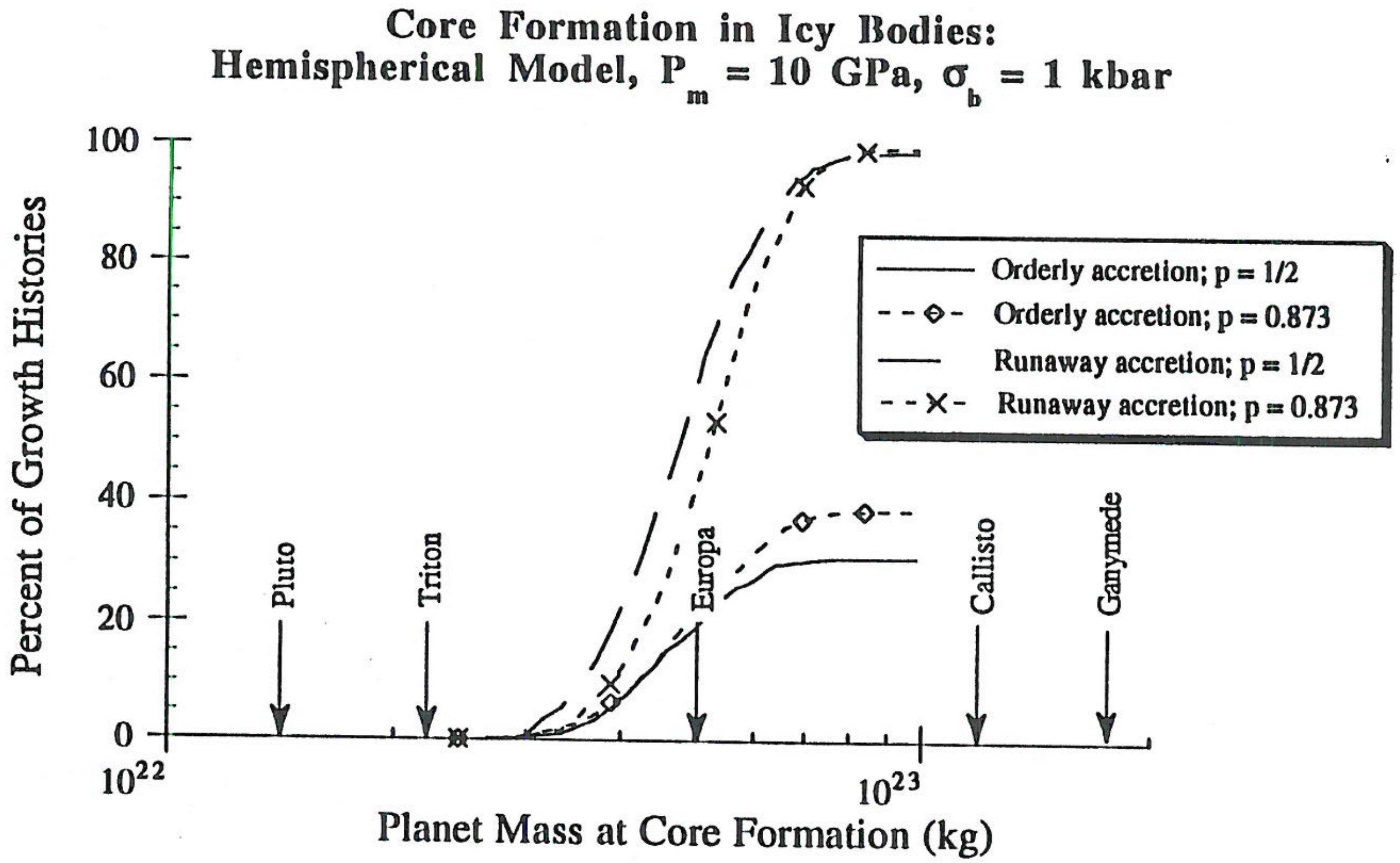
Core Formation in Icy Bodies:
Hemispherical Model, $P_m$ = 10 GPa, $\sigma_b$ = 1 kbar
100
80
60
40
20
0
$10^{22}$
$10^{23}$
Percent of Growth Histories
Planet Mass at Core Formation (kg)
Pluto
Triton
Europa
Callisto
Ganymede
Orderly accretion; p = 1/2
Orderly accretion; p = 0.873
Runaway accretion; p = 1/2
Runaway accretion; p = 0.873

Fig 6

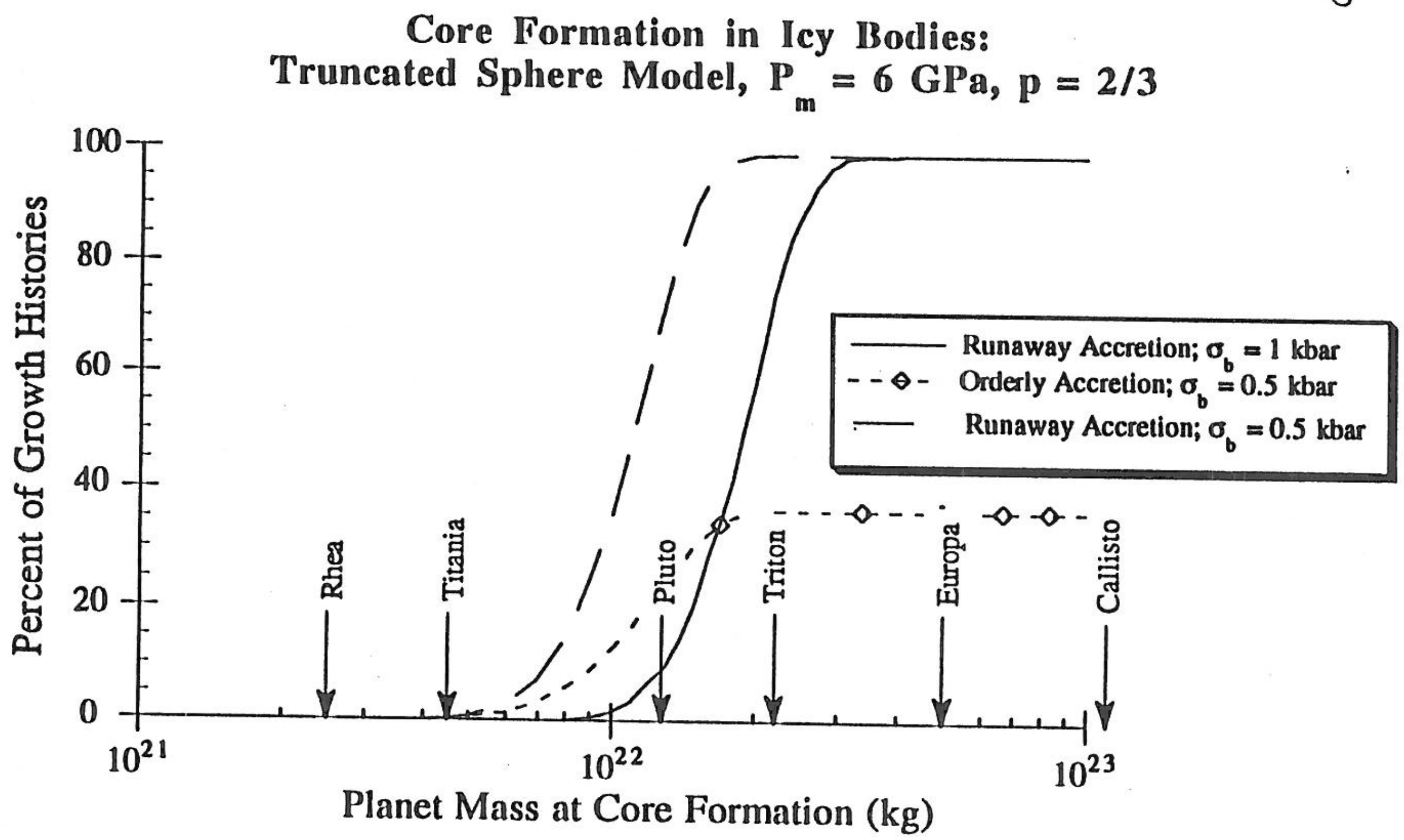
Core Formation in Icy Bodies:
Truncated Sphere Model, $P_m$ = 6 GPa, p = 2/3
Percent of Growth Histories
100
80
60
40
20
0
$10^{21}$
$10^{22}$
$10^{23}$
Planet Mass at Core Formation (kg)
Rhea
Titania
Pluto
Triton
Europa
Callisto
Runaway Accretion; $\sigma_b$ = 1 kbar
Orderly Accretion; $\sigma_b$ = 0.5 kbar
Runaway Accretion; $\sigma_b$ = 0.5 kbar

Fig 7

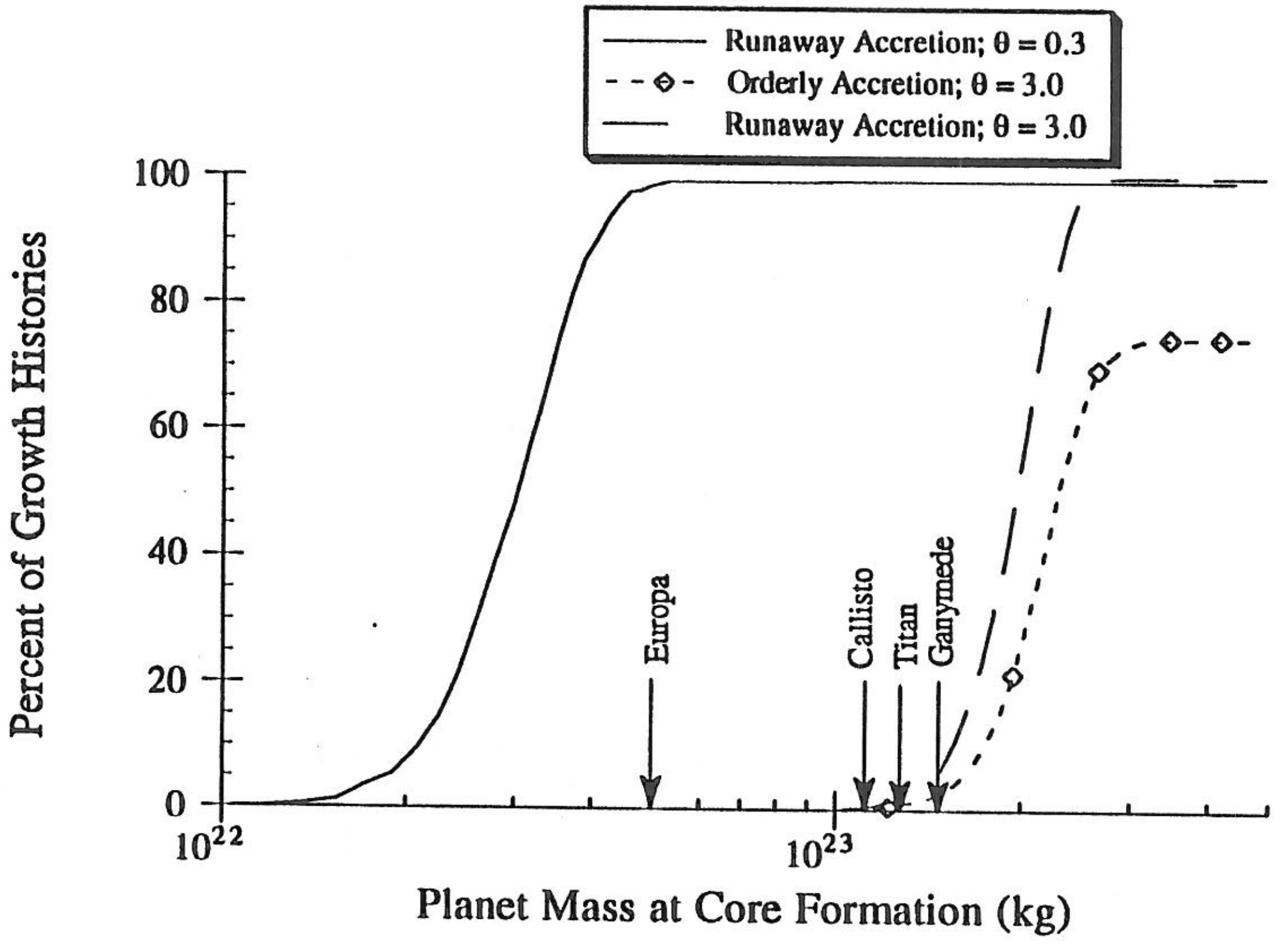

Runaway Accretion; θ = 0.3
Orderly Accretion; θ = 3.0
Runaway Accretion; θ = 3.0
100
80
60
40
20
0
Percent of Growth Histories
Europa
Callisto
Titan
Ganymede
10^22
10^23
Planet Mass at Core Formation (kg)

Fig 8

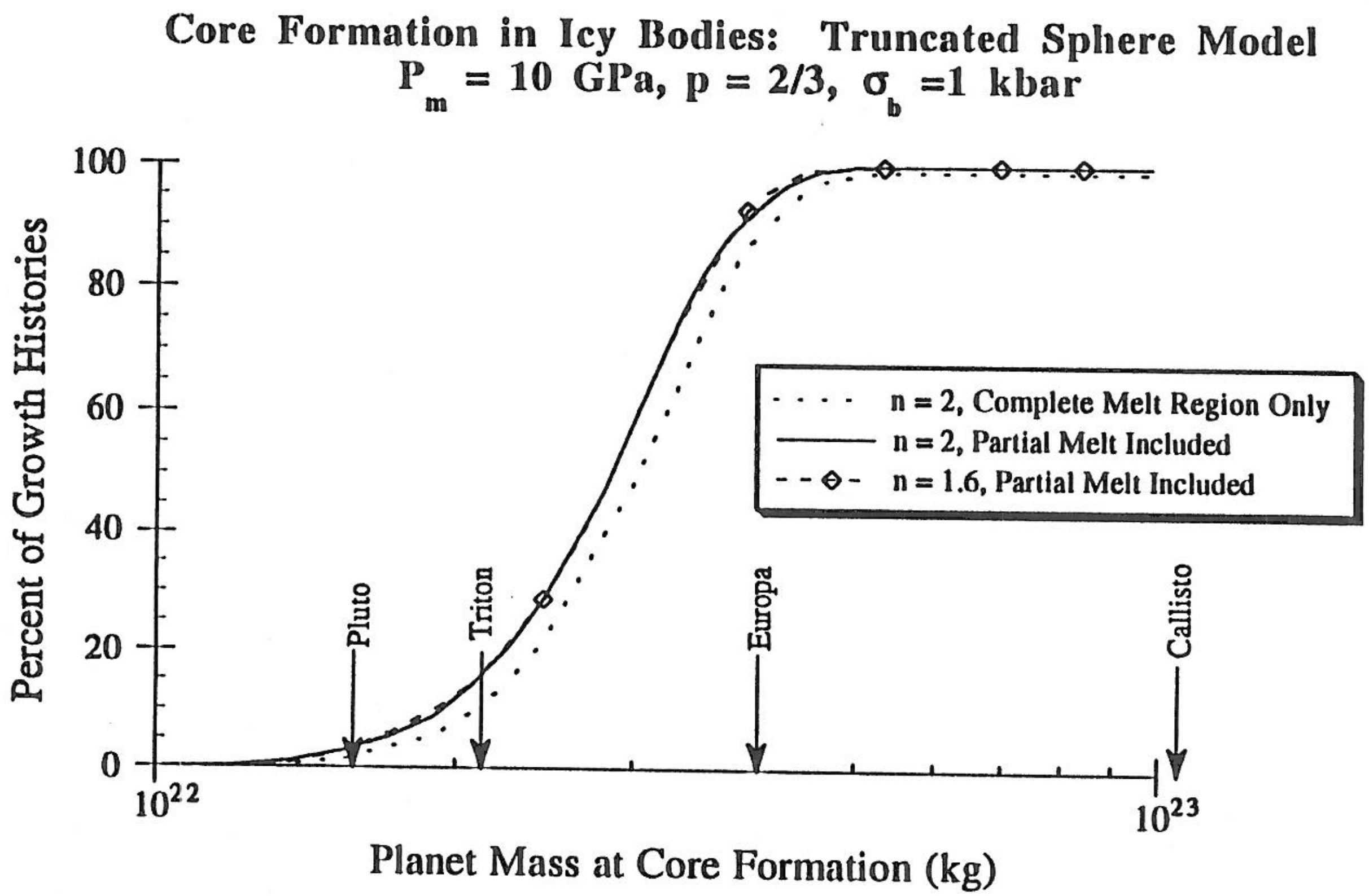
Core Formation in Icy Bodies: Truncated Sphere Model
$P_m$ = 10 GPa, p = 2/3, $\sigma_b$ =1 kbar
Percent of Growth Histories
100
80
60
40
20
0
$10^{22}$
$10^{23}$
Planet Mass at Core Formation (kg)
Pluto
Triton
Europa
Callisto
n = 2, Complete Melt Region Only
n = 2, Partial Melt Included
n = 1.6, Partial Melt Included

Fig 9

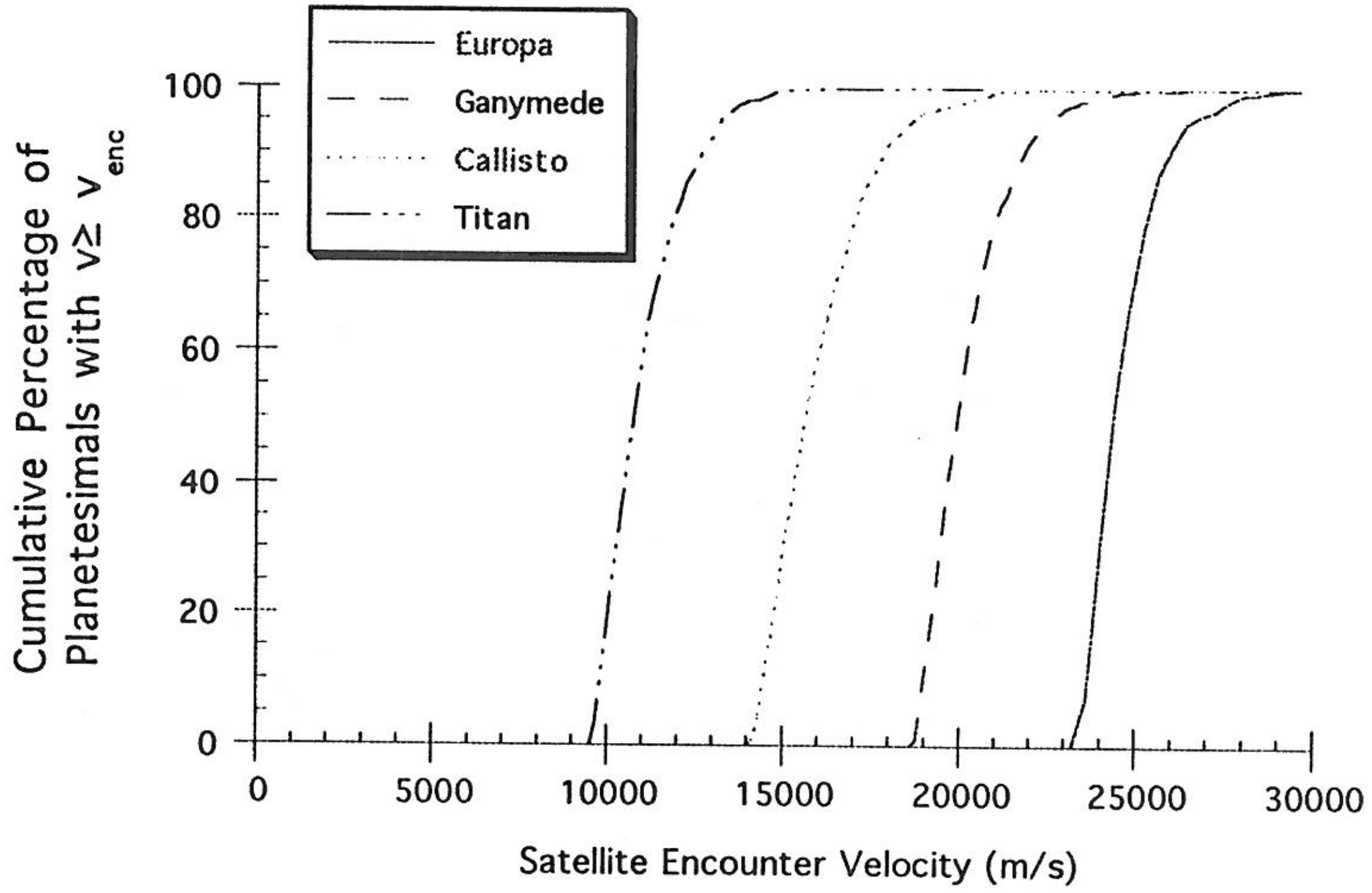

Europa
Ganymede
Callisto
Titan
100
80
60
40
20
0
0
5000
10000
15000
20000
25000
30000
Cumulative Percentage of Planetesimals with v≥ v_enc
Satellite Encounter Velocity (m/s)

Fig 10

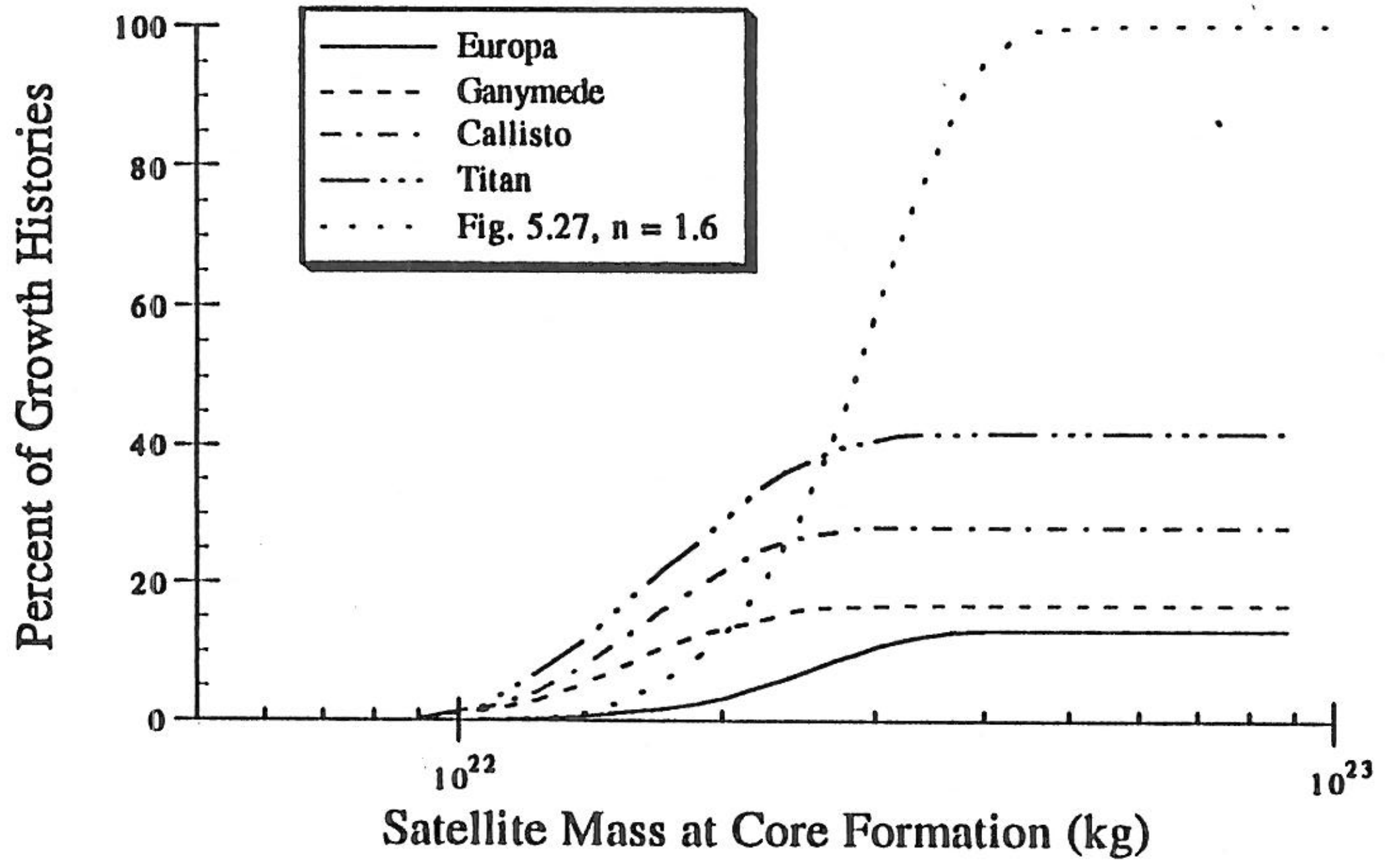

100
80
60
40
20
0
Percent of Growth Histories
Europa
Ganymede
Callisto
Titan
Fig. 5.27, n = 1.6
10^22
10^23
Satellite Mass at Core Formation (kg)

Fig 11

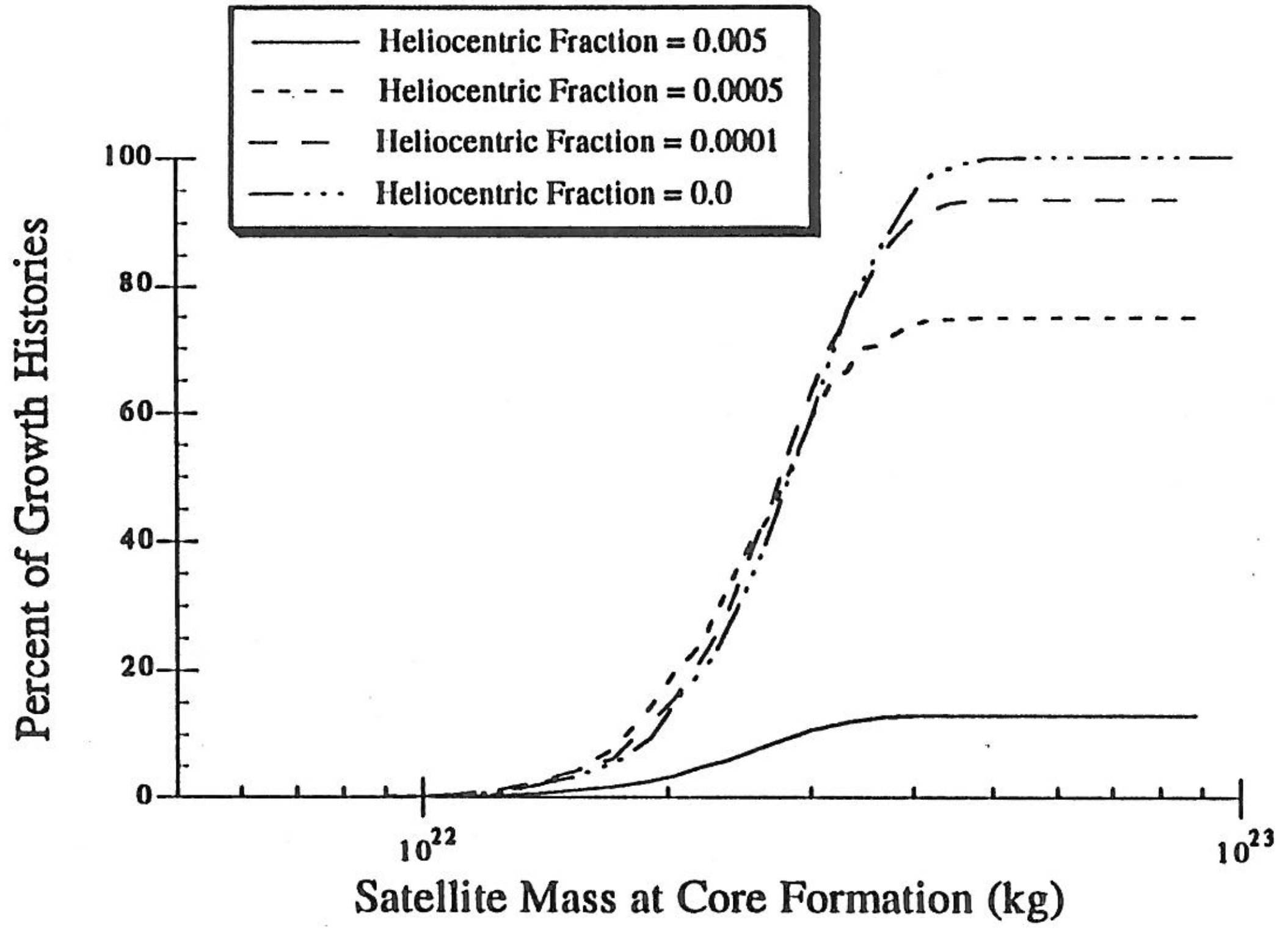

Heliocentric Fraction = 0.005
Heliocentric Fraction = 0.0005
Heliocentric Fraction = 0.0001
Heliocentric Fraction = 0.0
100
80
60
40
20
0
Percent of Growth Histories
$10^{22}$
$10^{23}$
Satellite Mass at Core Formation (kg)

Fig 12

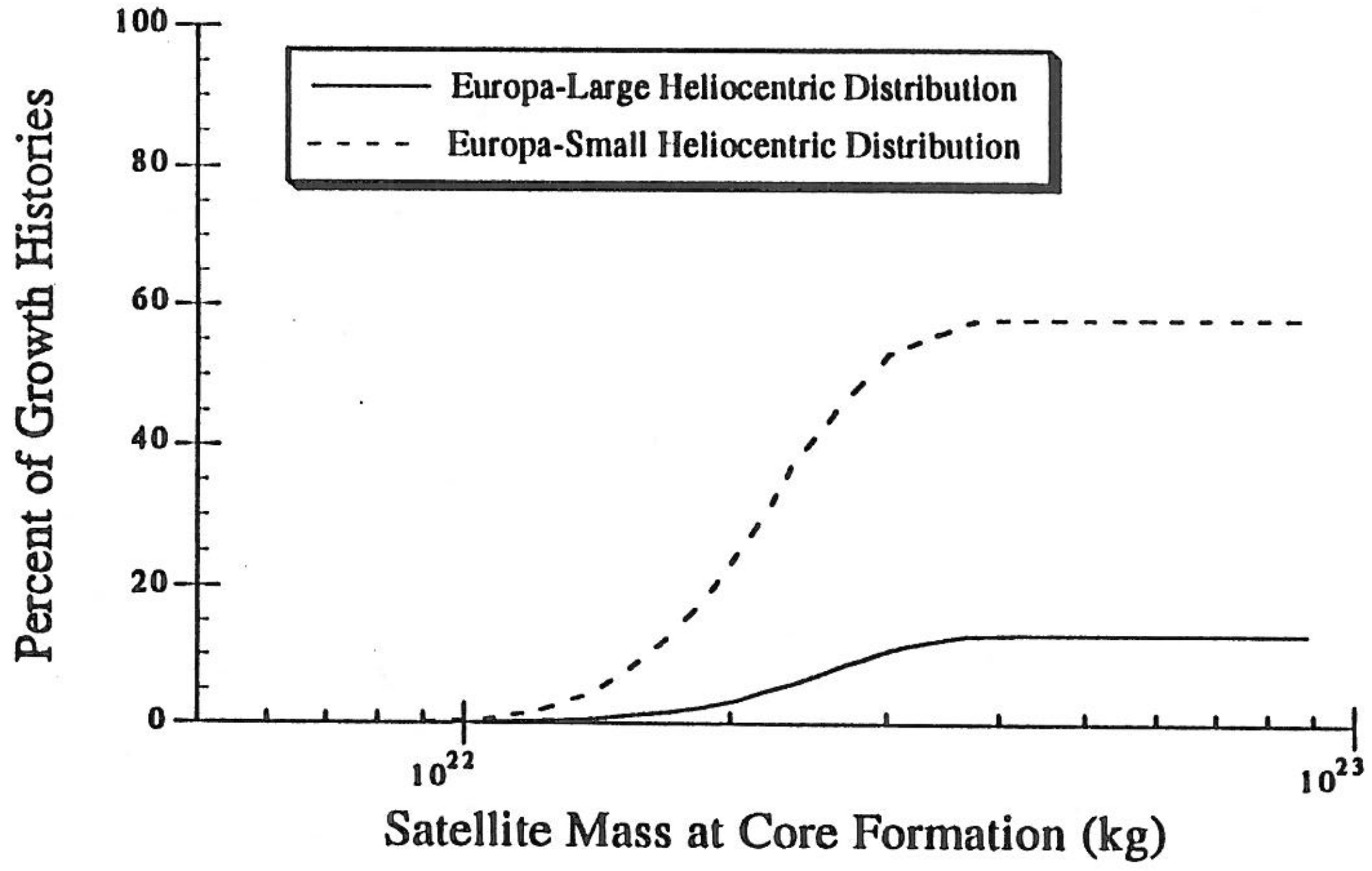

Europa-Large Heliocentric Distribution
Europa-Small Heliocentric Distribution
Percent of Growth Histories
100
80
60
40
20
0
$10^{22}$
$10^{23}$
Satellite Mass at Core Formation (kg)

Fig 13

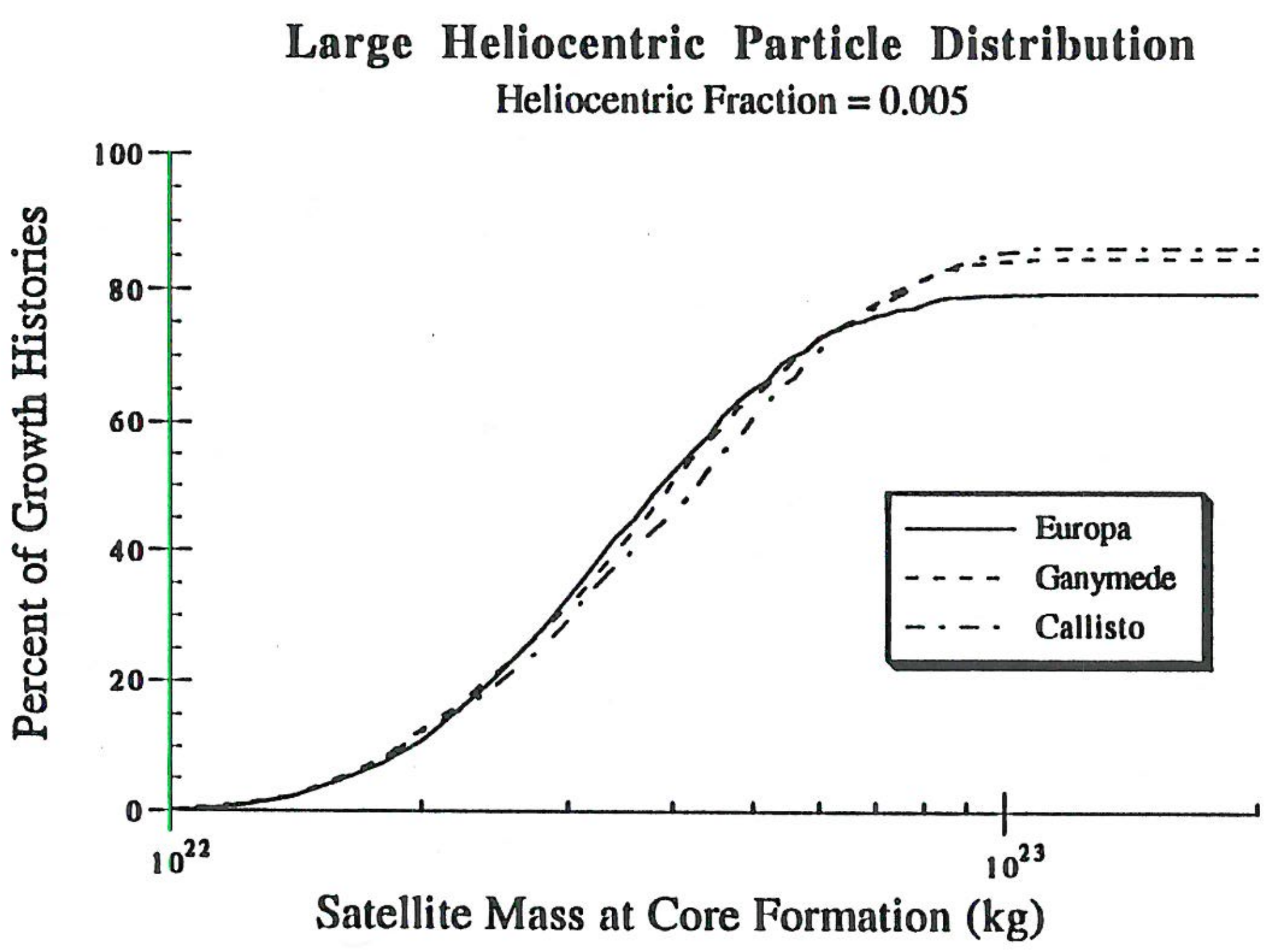
Large Heliocentric Particle Distribution
Heliocentric Fraction = 0.005
Percent of Growth Histories
100
80
60
40
20
0
$10^{22}$
$10^{23}$
Satellite Mass at Core Formation (kg)
Europa
Ganymede
Callisto

Fig 14

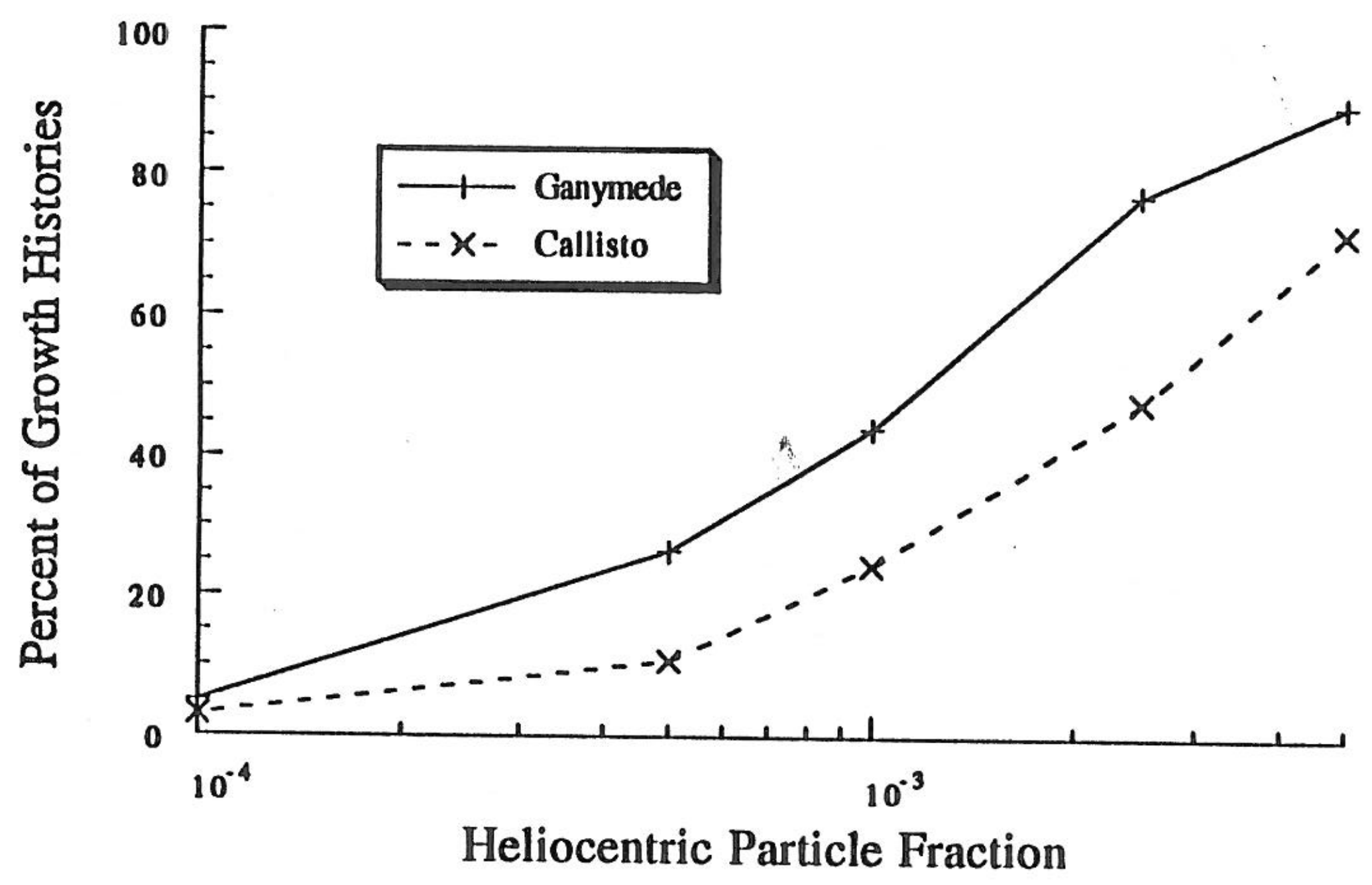

100
80
60
40
20
0
Percent of Growth Histories
Ganymede
Callisto
10-4
10-3
Heliocentric Particle Fraction

Fig. 15

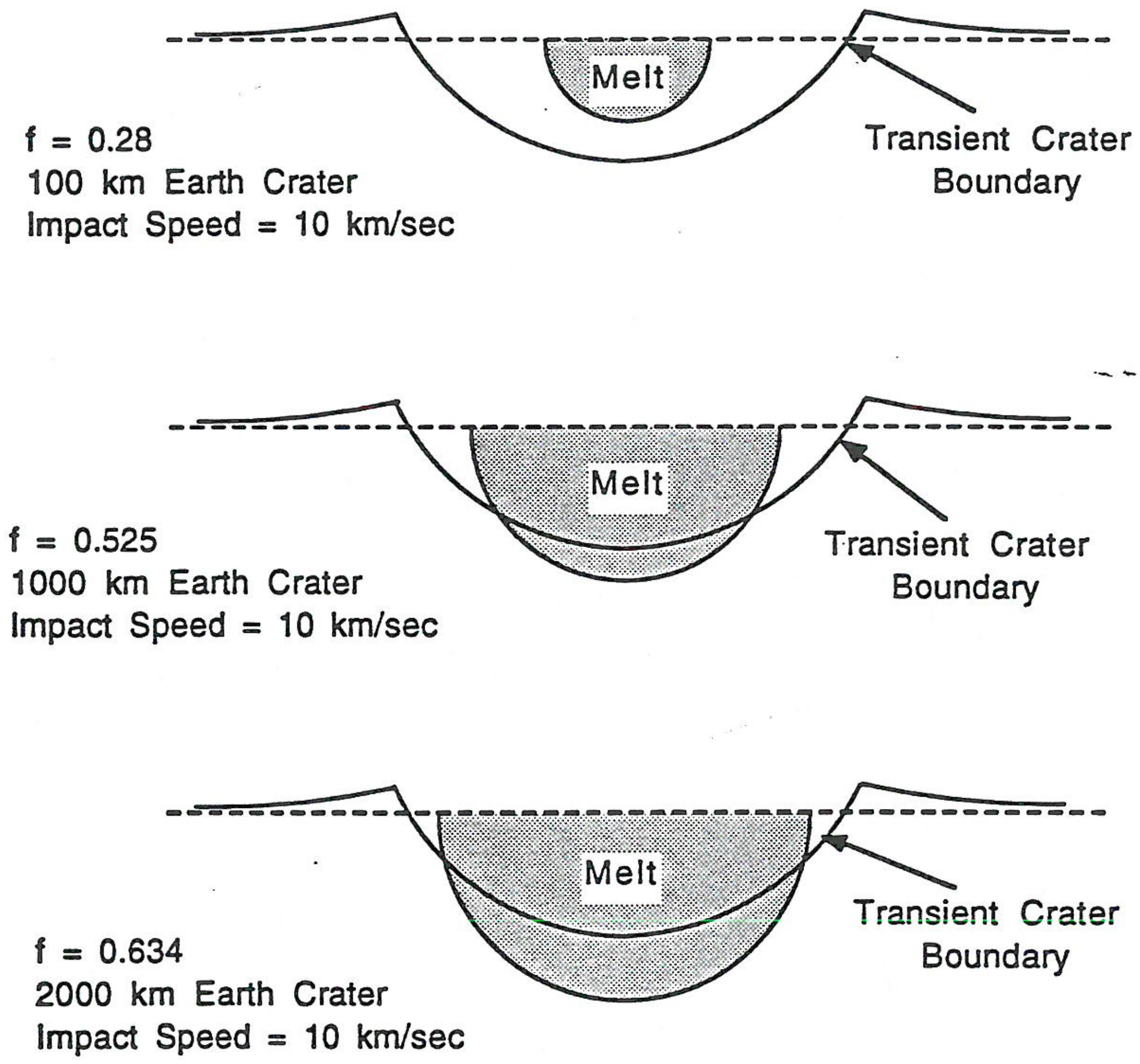
Melt
Transient Crater Boundary
f = 0.28
100 km Earth Crater
Impact Speed = 10 km/sec
Melt
Transient Crater Boundary
f = 0.525
1000 km Earth Crater
Impact Speed = 10 km/sec
Melt
Transient Crater Boundary
f = 0.634
2000 km Earth Crater
Impact Speed = 10 km/sec

Fig 16a

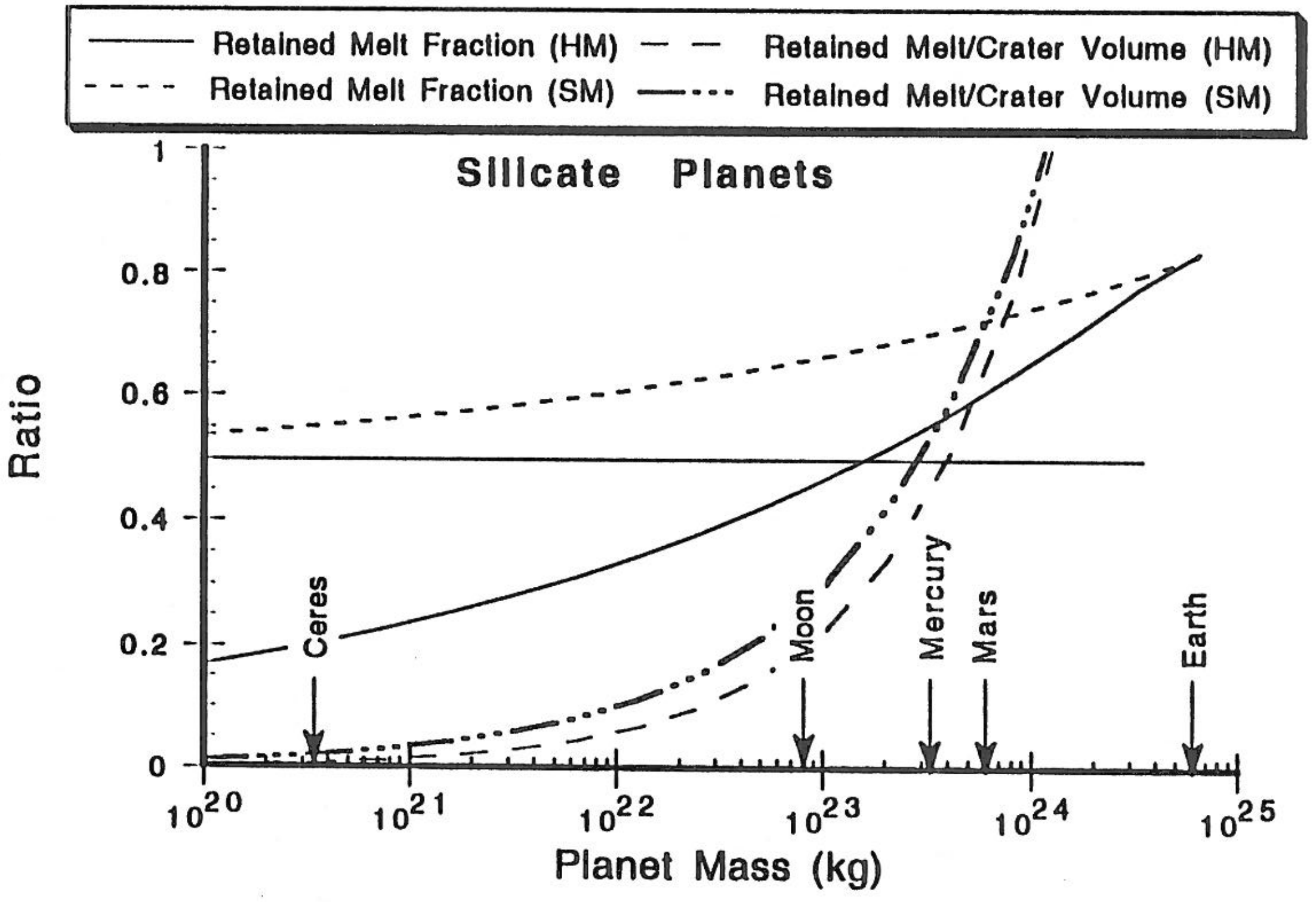

Retained Melt Fraction (HM)
Retained Melt Fraction (SM)
Retained Melt/Crater Volume (HM)
Retained Melt/Crater Volume (SM)
Silicate Planets
Ratio
1
0.8
0.6
0.4
0.2
0
Ceres
Moon
Mercury
Mars
Earth
$10^{20}$
$10^{21}$
$10^{22}$
$10^{23}$
$10^{24}$
$10^{25}$
Planet Mass (kg)

Fig 16b

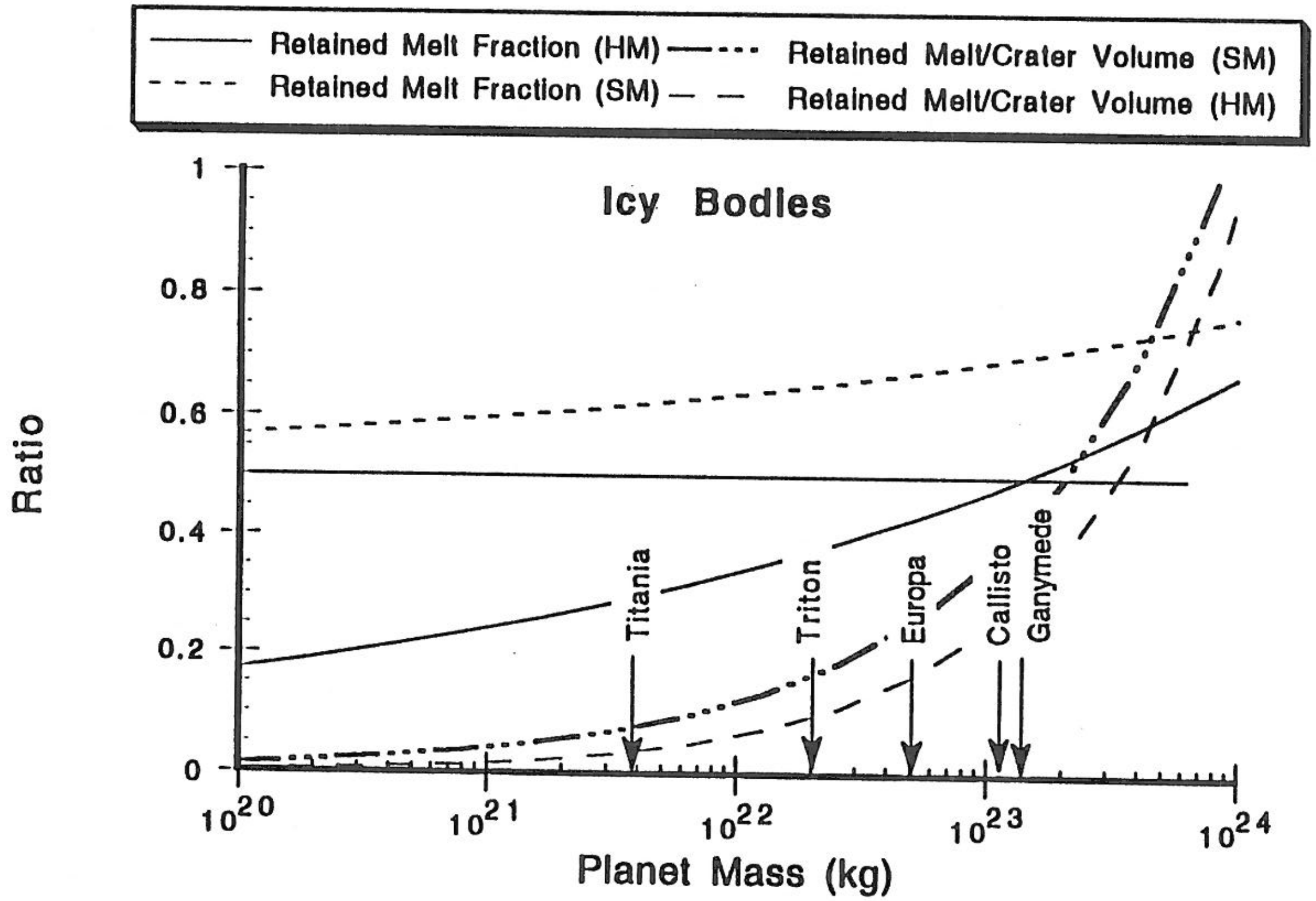
Retained Melt Fraction (HM)
Retained Melt Fraction (SM)
Retained Melt/Crater Volume (SM)
Retained Melt/Crater Volume (HM)
Icy Bodies
Ratio
1
0.8
0.6
0.4
0.2
0
Titania
Triton
Europa
Callisto
Ganymede
10^20
10^21
10^22
10^23
10^24
Planet Mass (kg)

Fig 17

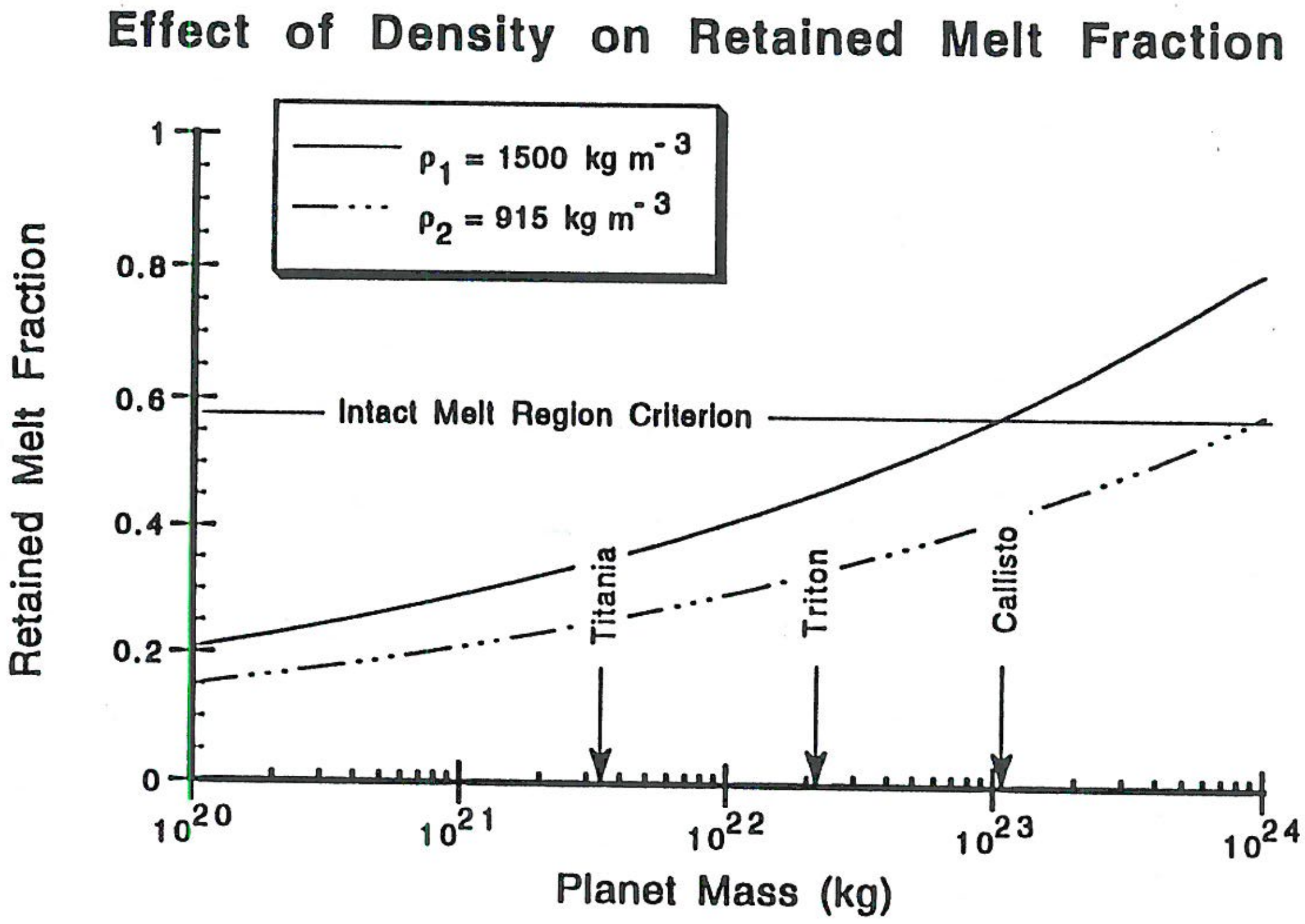

Effect of Density on Retained Melt Fraction
ρ1 = 1500 kg m-3
ρ2 = 915 kg m-3
Intact Melt Region Criterion
Titania
Triton
Callisto
Retained Melt Fraction
0
0.2
0.4
0.6
0.8
1
10^20
10^21
10^22
10^23
10^24
Planet Mass (kg)

Fig A1a

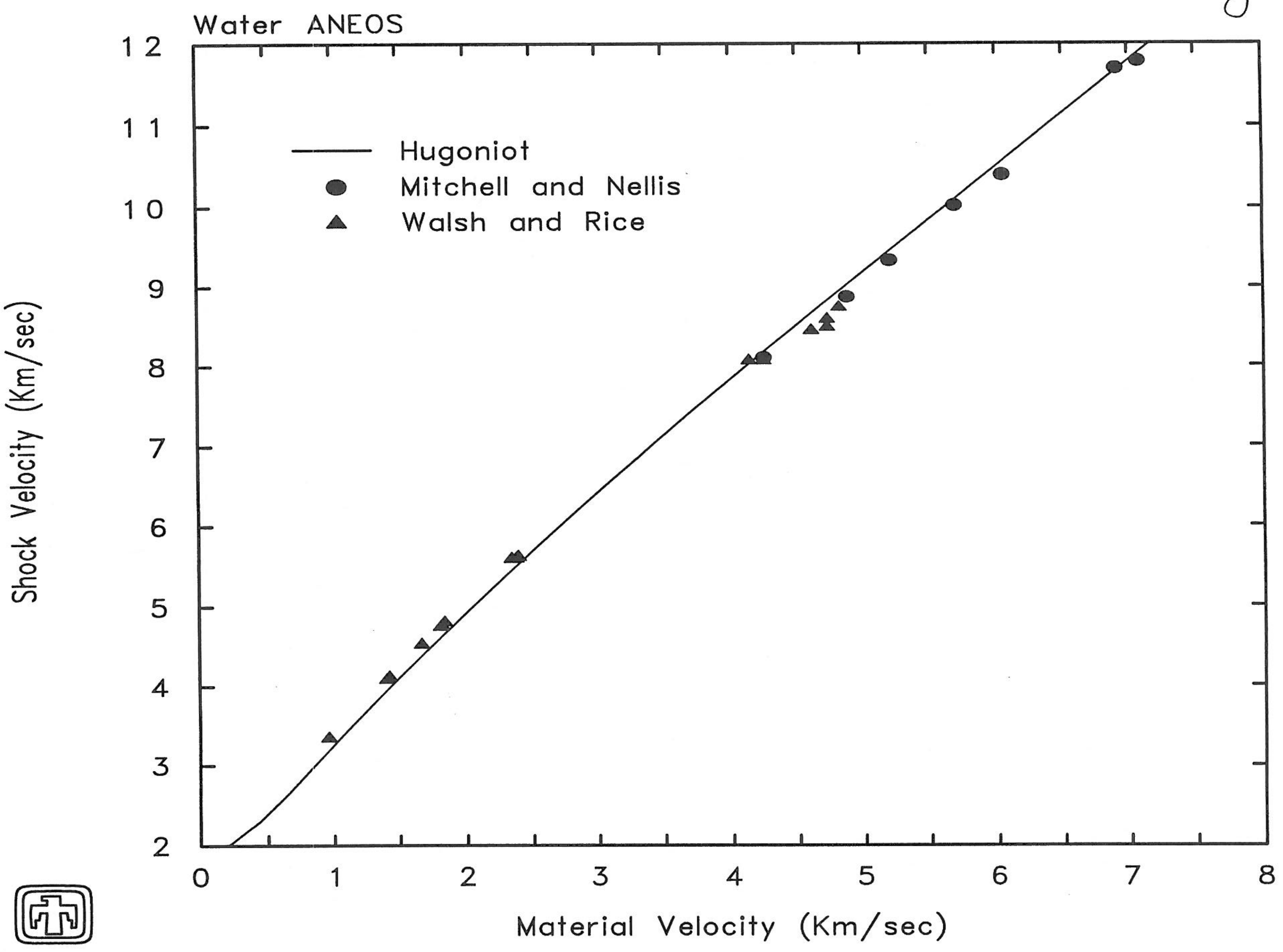

Water ANEOS
Hugoniot
Mitchell and Nellis
Walsh and Rice
Shock Velocity (Km/sec)
Material Velocity (Km/sec)
2
3
4
5
6
7
8
9
10
11
12
0
1
2
3
4
5
6
7
8

Fig A1b

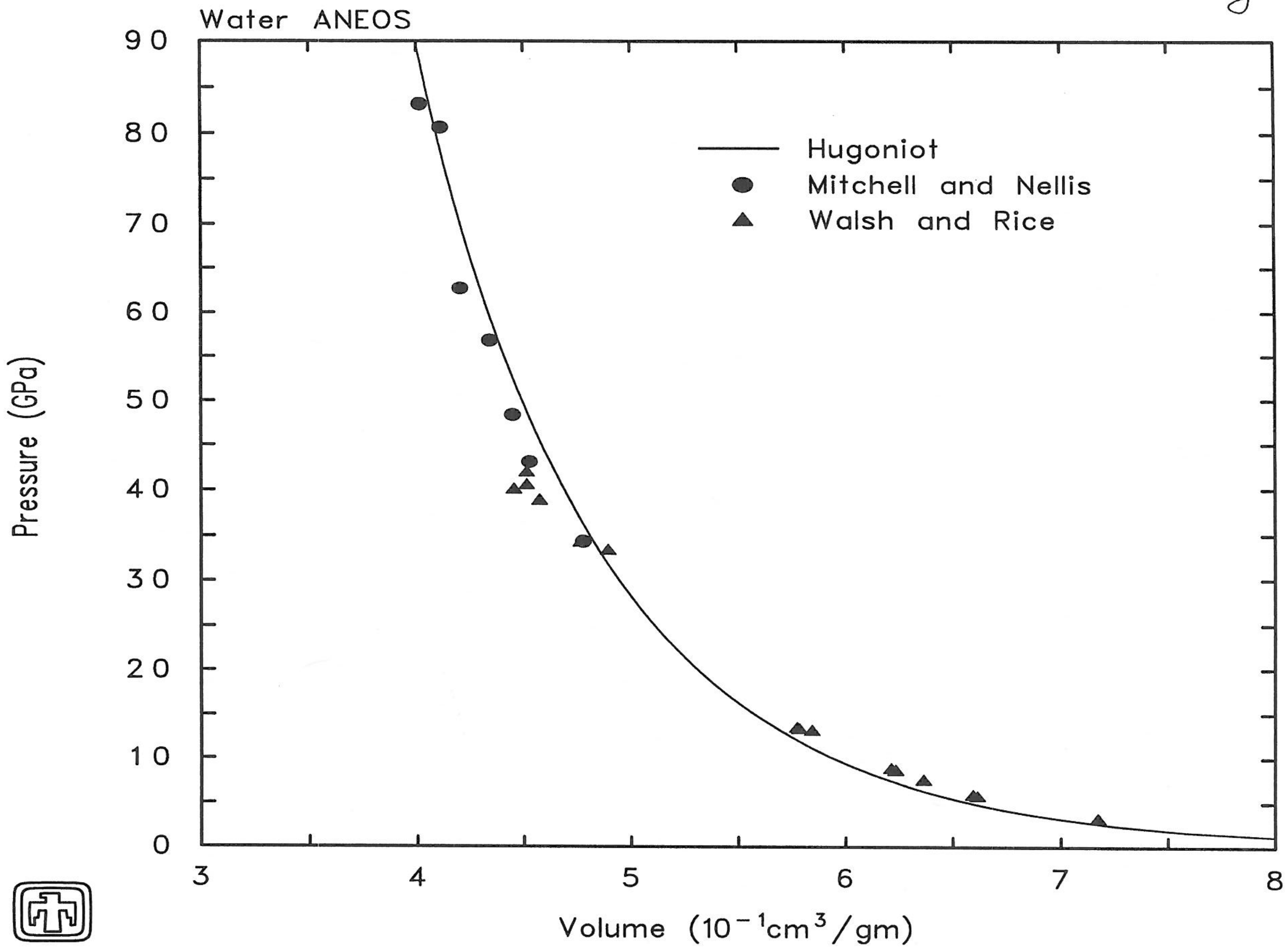

Water ANEOS
Hugoniot
Mitchell and Nellis
Walsh and Rice
Pressure (GPa)
90
80
70
60
50
40
30
20
10
0
3
4
5
6
7
8
Volume ($10^{-1}cm^3/gm$)

Fig. A2

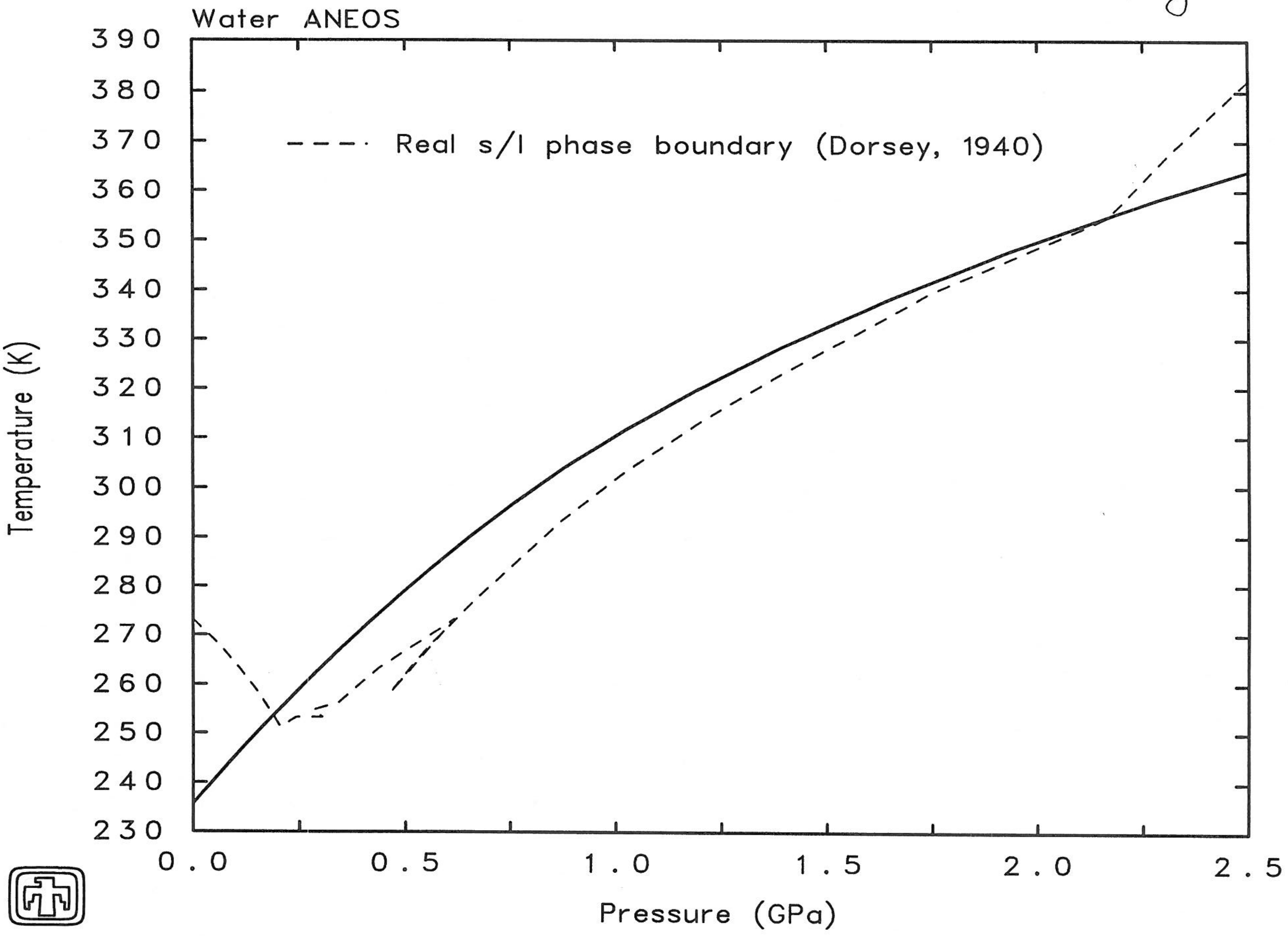

Water ANEOS
Real s/l phase boundary (Dorsey, 1940)
Temperature (K)
390
380
370
360
350
340
330
320
310
300
290
280
270
260
250
240
230
0.0
0.5
1.0
1.5
2.0
2.5
Pressure (GPa)

Fig. 13

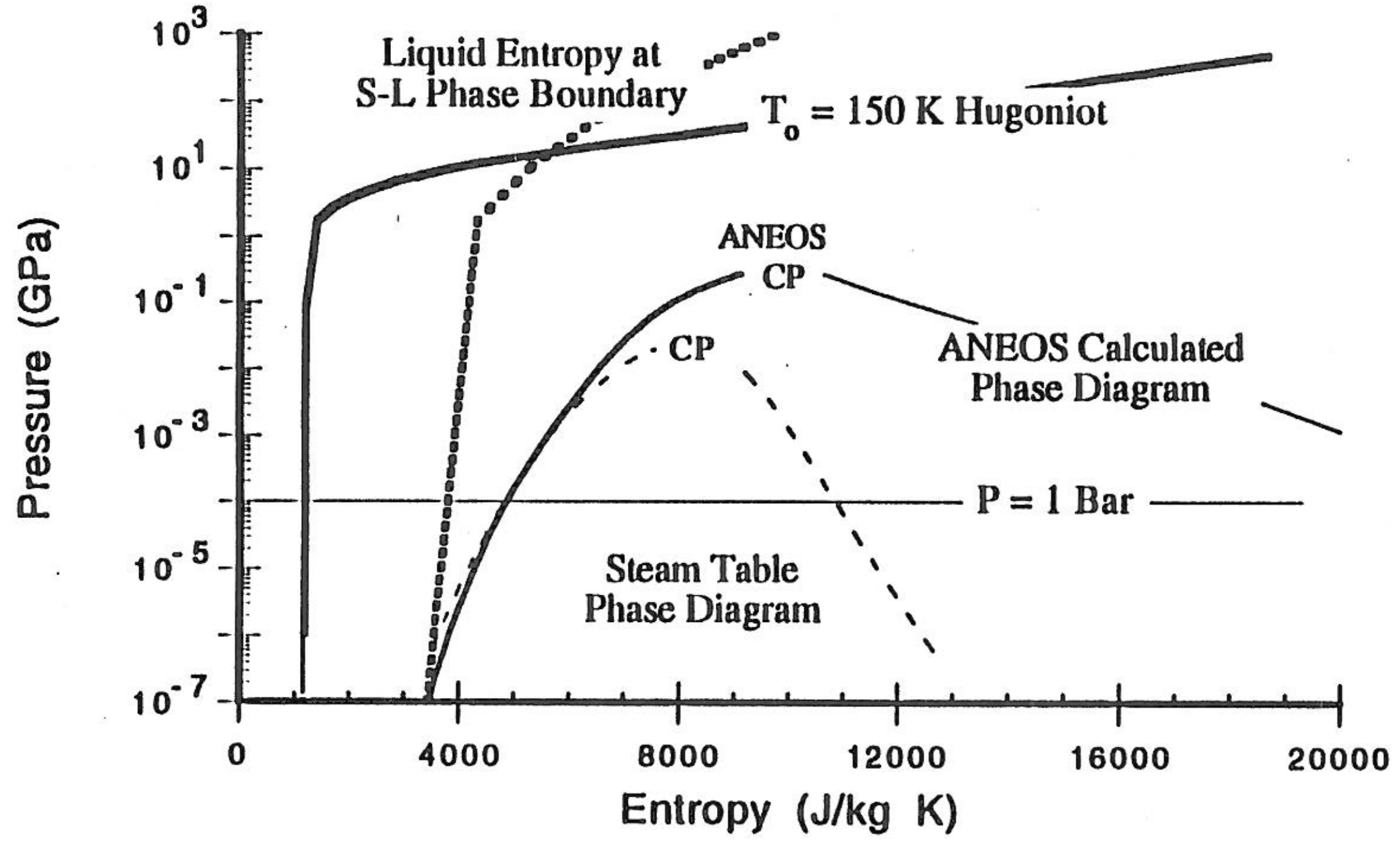

Liquid Entropy at
S-L Phase Boundary
T_o = 150 K Hugoniot
ANEOS
CP
CP
ANEOS Calculated
Phase Diagram
P = 1 Bar
Steam Table
Phase Diagram
Pressure (GPa)
10^3
10^1
10^-1
10^-3
10^-5
10^-7
0
4000
8000
12000
16000
20000
Entropy (J/kg K)